\documentclass[traditabstract]{aa}

\usepackage[dvips]{graphicx}
\usepackage{times}  
\usepackage{natbib}
\usepackage{lscape}
\usepackage{aalongtable,ltcaption}

\begin{document}

\title{The Hamburg/ESO R-process Enhanced Star survey (HERES) 
\thanks{Based on observations collected at the European Southern Observatory, Paranal, Chile (Proposal Number 68.B-0320).} 
} 

\subtitle{VII. Thorium abundances in metal-poor stars}

\author{J. Ren\inst{1,}\inst{2,}\inst{3} 
\and N. Christlieb\inst{2}
\and G. Zhao\inst{1,}\inst{4}
\institute{The School of Space Science and Physics, Shandong University at Weihai, Wenhua Xilu 180, 264209 Weihai, 
Shandong, China\\
\email{renjing@bao.ac.cn,gzhao@nao.cas.cn}
\and Zentrum f$\ddot{\mathrm{u}}$r Astronomy der Universit$\ddot{\mathrm{a}}$t Heidelberg, Landessternwarte, 
K$\ddot{\mathrm{o}}$nigstuhl 12, D-69117 Heidelberg, Germany \\
\email{N.Christlieb@lsw.uni-heidelberg.de}
\and Department of Physics and Astronomy, Uppsala University, Box 516, 751 20 Uppsala, Sweden 
\and Key Laboratory of Optical Astronomy, National Astronomical Observatories, CAS, 20A Datun Road, 100012 Chaoyang 
District, Beijing, China}
}

\date{Received  / Accepted }

\abstract{We report thorium abundances for 77 metal-poor stars in the metallicity range of $-3.5<\mathrm{[Fe/H]}<-1.0$, 
based on ``snapshot'' spectra obtained with VLT-UT2/UVES during the HERES Survey. We were able to determine the thorium 
abundances with better than $1\sigma$ confidence for 17 stars, while for 60 stars we derived upper limits. For five stars 
common with previous studies, our results were in good agreement with the literature results. The thorium abundances 
span a wide range of about 4.0 dex, and scatter exists in the distribution of log(Th/Eu) ratios for lower metallicity stars, 
supporting previous studies suggesting the r-process is not universal. We derived ages from the log(Th/Eu) ratios for 12 stars, 
resulting in large scattered ages, and two stars with significant enhancement of Th relative to Eu are found, indicating 
the ``actinide boost'' does not seem to be a rare phenomenon and thus highlighting the risk in using log(Th/Eu) to derive stellar 
ages.
\keywords{Stars: abundances -- Stars: population II -- Galaxy: abundances -- Galaxy: evolution -- Galaxy: halo }}
\maketitle
%%%%%%%%%%%%%%%%%%%%%%%%%%%%%%%%%%%%%%%%%%%%%%%%%%%%%%%%%%%%%%%%%%%%%%%%%%%%%%%%%%%%%%%%%%%%%%%%%%%%%%%%%%%%%%%%%
\section{Introduction}
The observed chemical abundances of metal-poor stars provide rich information on star formation and nucleosynthesis in the 
early stages of evolution of our galaxy and other galaxies. In particular, the study of heavy elements beyond the iron 
group in metal-poor stars have greatly improved the understanding of the neutron-capture nucleosynthesis processes in 
the early Universe, e.g., placing constraints on the astrophysical site(s) of r-process. Neutron-capture elements in 
r-process enhanced metal-poor stars have been reported to match the scaled solar r-process abundance pattern at least 
in the range of $\mathrm{Z}=52\sim76$ \citep[see e.g.,][]{cowan:02,sneden:03,honda:04}. In contrast, lighter elements 
($\mathrm{Z}<50$) abundances deviate significantly from the Solar-pattern, indicating the existence of different initial 
production ratios for the r-process elements \citep{sneden:00,sneden:03,honda:04}. In old, metal-poor stars, thorium 
and uranium with relatively long half-life are the only heavier r-process elements we could observe today, and offer 
a way to better understand the heaviest products in r-process nucleosynthesis \citep{roederer:09}. 

The detection of the radioactive-decay elements in r-process enhanced metal-poor stars allowed a new approach to derive 
the age of the oldest stars, by means of comparing the observed ratios of radioactive decay elements over stable elements
 with the corresponding initial values at the time when the star was born\footnote{Accurately speaking, it was the time 
 when the radioactive decay elements were produced, not necessarily to be the same as when the star was born.}, 
 hence providing a lower limit on the age of the universe. The actinide thorium is one such radioactive species. 
 It is produced in the r-process, and has been widely used, particularly log(Th/Eu), for cosmochronometry to estimate 
 stellar ages \citep[e.g.,][]{sneden:96,hill:02,honda:04,frebel:07,hayek:09}. Some authors have argued that log(Th/Eu) 
 is not a reliable chronometer to derive stellar ages, because of the relatively large uncertainties 
 \citep[e.g.,][]{goriely:99,goriely:01,honda:04,frebel:07}. Due to its longer half-life of $14.05$ Gyr as compared 
 with, say, $4.5$\,Gyr of uranium, some authors have suggested that the log(U/Th) pair would be a better chronometer, 
 because of their close atomic number and thus relatively small nuclear physics uncertainties 
 \citep{cayrel:01,wanajo:02,beers:05}. Additionally, because Pb and Bi mostly originate from $\alpha$-decay of the 
 Th and U isotopes, the measurement of Pb or Bi abundances can offer a consistency check on the calculated initial 
 abundances for this pair \citep{cowan:99}. However, it is very difficult to detect U lines in stellar spectra, due 
 to the blending of the weak U lines (e.g., 3859 \AA~) and the relatively low amount of uranium in an old star with 
 age of $\sim12-15$\,Gyr \citep[see][]{plez:04,roederer:09}. Alternatively, Hf behaves similarly to third-peak elements 
 and the log(Th/Hf) ratio has been suggested by \citet{kratz:07} as a promising tool for chronometer studies.    

Many efforts have been made to determine thorium abundances in metal-poor stars using high resolution spectra. 
So far, more than 30 metal-poor stars ($\mathrm{[Fe/H]}<-1$) have been reported with thorium abundances in previous 
studies, and the error is typically of order of $0.15$ dex. It appears that among the r-process enhanced stars with 
measured Th or other actinide elements (e.g.,U), except for an actinide normal group, which means no obvious enhancement 
of actinide element abundances with respect to the scaled solar r-process pattern, an actinide boost group 
(e.g., CS31082-001, HE1219-0312, CS30306-132) also came to be known, although the former group seems to be more common 
among r-process enhanced metal-poor stars (e.g., CS22892-052, CS29497-004). This imples that for elements in the 
range of $\mathrm{Z}\ge90$, significantly different chemical yields might be produced due to the various conditions 
of the star formation regions \citep{hill:02,roederer:09,mashonkina:10}.

To obtain accurate age estimates and to explore r-process nucleosynthesis in metal-poor stars in detail, high-precision 
Th abundances are needed. Further, large samples of stars with Th abundances are also important to study the 
distribution of thorium abundances in metal-poor stars. The HERES survey offers a good opportunity to perform 
such a study. More than 22 elemental abundances (not including Th) for the sample stars have been reported in 
previously published papers in this series\citep[e.g.,][]{barklem:05,zhang:10}, and thorium abundances have been 
studied furtherly with better quality spectra in a few stars \citep[][]{hayek:09,mashonkina:10}. 

Here we report thorium abundances for 77 stars from the sample of \citet{barklem:05} (hereafter Paper II), using a 
modified version of the analysis method described in that paper. A brief description of the sample is given in 
sect. \ref{sect:obs}. The abundance analysis is described in sect. \ref{sect:anay}. In sect. \ref{sect:thorium}, 
thorium abundances are presented and compared with previous works, and the implications for stellar age estimates 
are discussed. Section \ref{sect:conclusion} presents the conclusions.
%%%%%%%%%%%%%%%%%%%%%%%%%%%%%%%%%%%%%%%%%%%%%%%%%%%%%%%%%%%%%%%%%%%%%%%%%%%%%%%%%%%%%%%%%%%%%%%%%%%%%%%%%%%%%%%%%%%%%%%%%%%%%%%
\section{The sample}
\label{sect:obs}
This work is based on the moderately high-resolution ``snapshot'' spectra of 253 HERES stars. Readers are referred 
to \citet{christlieb:04} and Paper II for detailed information on the observations and the sample selection.  
For convenience, we repeat the most important basic information on the sample. The spectra 
($R\sim20000$, $\lambda=3760-4980$\,\AA~, and typical $S/N\sim30$ to $50$) were obtained during the HERES survey 
with ESO-VLT2/UVES. A total of 373 spectra were observed, which was reduced to 253 stars in Paper II when stars 
were removed for various reasons, most importantly due to strong molecular carbon features leading to significant 
blending.  Much of the remainder of the sample has been analysed by \citet{lucatello:06}.  These 253 stars are 
the starting point for our analysis; however, reasonable estimates of Th abundances or even upper limits could 
not be obtained for all stars for reasons that will be discussed further below.
%%%%%%%%%%%%%%%%%%%%%%%%%%%%%%%%%%%%%%%%%%%%%%%%%%%%%%%%%%%%%%%%%%%%%%%%%%%%%%%%%%%%%%%%%%%%%%%%%%%%%%%%%%%%%%%%%%%%%%%%%%%%%%%%
\section{Abundance analysis}
\label{sect:anay}
We derive abundances from the \ion{Th}{II} 4019.12~\AA~ line, the only Th line strong enough to be detected in our spectra.  
This line is unfortunately blended, including blends of \element[ ][13]{CH}.  The line has been analysed using the  
automated spectrum analysis code based on SME \citep{valenti:96} described in Paper II and used in \citet{jonsell:06} 
(hereafter Paper III).  The stellar parameters and abundances of other elements, particularly those giving rise to 
blends with the Th line, were adopted directly from Paper II.  In order to be able to model the \element[ ][13]{CH} 
blends, some modifications were made to the code to enable estimates of the \element[ ][12]{C}/ \element[ ][13]{C} 
ratio to be obtained. This ratio has been determined from isolated \element[ ][13]{CH} features between 4210 and 4225~\AA~ 
and at 4370~\AA, as shown in Paper III, though now in an automated manner. 

The employed line data is shown in Table~\ref{tab:linelist}, and is essentially that of \citet{johnson:01}.  
The $f$ value for the Th II line has been updated (a change of +0.05 dex) and the wavelengths and excitation potentials 
have been negligibly changed in some cases based on VALD values \citep{vald2a}.  All blends within 0.5~\AA\ either 
side of the Th line in the list of \citet{johnson:01} are included, except a line of U and and a line of V, 
which we are unable to model since we do not have abundances for these elements; their contributions are expected 
to be negligible in any case.  A window 0.3~\AA\ wide centred on the Th line is used for the fitting.
\begin{table}
\centering
\caption{Line list.}
\label{tab:linelist}
\begin{tabular}{rrrrc}
\hline \hline
Species & $\lambda $ & $\xi $ & $\log gf $ & Refs. \\
        & [{\AA}]    & [eV]   &            &       \\
\hline
\ion{Th}{II}  & 4019.129  & 0.000  & $-0.228$ &      1 \\
&&&&\\
\ion{\element[ ][13]{CH}}{I}  & 4019.000  & 0.460  & $-1.163$ &      2 \\
\ion{\element[ ][13]{CH}}{I}  & 4019.170  & 0.460  & $-1.137$ &      2 \\
&&&&\\
\ion{Fe}{I}  & 4019.050  & 2.608  & $-2.780$ &      3 \\
&&&&\\    
\ion{Co}{I}  & 4019.110  & 2.280  & $-3.287$ &      2 \\
\ion{Co}{I}  & 4019.118  & 2.280  & $-3.173$ &      \\
\ion{Co}{I}  & 4019.120  & 2.280  & $-3.876$ &      \\
\ion{Co}{I}  & 4019.125  & 2.280  & $-3.298$ &      \\
\ion{Co}{I}  & 4019.125  & 2.280  & $-3.492$ &      \\
\ion{Co}{I}  & 4019.134  & 2.280  & $-3.287$ &      \\
\ion{Co}{I}  & 4019.135  & 2.280  & $-3.474$ &      \\ 
\ion{Co}{I}  & 4019.138  & 2.280  & $-3.173$ &      \\
\ion{Co}{I}  & 4019.140  & 2.280  & $-3.298$ &      \\
&&&&\\ 
\ion{Co}{I}  & 4019.272  & 0.580  & $-3.480$ &      2 \\
\ion{Co}{I}  & 4019.281  & 0.580  & $-3.470$ &      \\
\ion{Co}{I}  & 4019.294  & 0.580  & $-3.220$ &      \\
\ion{Co}{I}  & 4019.296  & 0.580  & $-3.330$ &      \\
\ion{Co}{I}  & 4019.322  & 0.580  & $-4.090$ &      \\
\ion{Co}{I}  & 4019.332  & 0.580  & $-4.040$ &      \\
&&&&\\ 
\ion{Ni}{I}  & 4019.058  & 1.935  & $-3.174$ &      2 \\
&&&&\\
\ion{Ce}{II}  & 4019.057  & 1.014  & $ 0.093$ &      2 \\
&&&&\\
\ion{Nd}{II}  & 4018.836  & 0.060  & $-0.880$ &      2 \\
\hline
\end{tabular}
\tablebib{(1) - \citet{Nilssonetal:2002}, (2) - \citet{johnson:01} (3) - \citet{vald2a}}
\end{table}

In Paper II we required at least one line to be detected at the 3$\sigma$ confidence level to claim an elemental detection.  
Such a high threshold would result in a very small number of detections given the weakness of the single Th line we used 
and the quality of our spectra.  Here we reduce this requirement to 1$\sigma$, though with some additional constraints. 
First, the significance of detection was caculated from the fit without blending, 
i.e., $m' = m \times (1-d_\mathrm{b}/d_\mathrm{fit})$, where $d_\mathrm{b}$ and $d_\mathrm{fit}$ are the line depths 
of the blendings (without thorium line) and the whole fit respectively; $m$ is the detection level of the 
the whole fit. Second, all lines calculated to have $m$ above 
the 1$\sigma$ detection level were subjected to manual inspection and adjustment adopted for some stars, usually due 
to noise or severe blending casting significant doubt on its validity. Finally, except for several stars raised to 
detection level considering their strong line strengths, for all other stars with detections $m'$ below the 1$\sigma$ 
confidence level and those rejected during manual inspection, we calculated 1$\sigma$ upper limits, but rejecting 
those with uncertain upper limits, i.e., blendings occupying more than $80\%$ of the fit or one of the fits not converged.    

The errors in the abundances could in principle be estimated using the methods of Paper II, but is complicated by 
the additional uncertainties from blending.    In Paper II we found relative and absolute errors of $\sim0.18$ 
and $\sim0.25$~dex, respectively, for species with similar atomic structure to Th II, and thus lines with similar 
sensitivities to stellar parameters, e.g. Nd II, Sm II, Eu II.  These cases have more and stronger lines and no 
significant blending issues.  Accounting for these extra sources of error we estimate relative and absolute errors 
in the Th abundances of $\sim0.25$ and $\sim0.3$~dex, respectively. Similarly, by comparison with errors in Nd/Eu etc., 
an error of $\sim0.25$ dex is estimated for the abundance ratio log(Th/Eu). 

However, it should be noted that, in our analysis, we have adopted the detection significance of $1\sigma$, which means 
the statistic uncertainty will dominate the general error. 
Thus, we value any interesting results from our analysis, but are very cautious to give any conclusion based on our results.
%%%%%%%%%%%%%%%%%%%%%%%%%%%%%%%%%%%%%%%%%%%%%%%%%%%%%%%%%%%%%%%%%%%%%%%%%%%%%%%%%%%%%%%%%%%%%%%%%%%%%%%%%%%%%%%%%%%%%%%%%%%%%%
\section{Results}
\label{sect:thorium}
The detected thorium abundances for stars classified into different subclasses are tabulated in Table ~\ref{tab:thdetect}. 
In Table \ref{tab:result} (online only), stars with upper limits are included, together with the atmospheric parameters and 
some other abundances of interest from Paper II.  The confidence level of the detection $m\sigma$ and $m'\sigma$, computed 
as described in section 3.4 of Paper II and in section \ref{sect:anay} of this paper, are also tabulated. The errors are 
not listed since they are either the same for all stars in the cases involving Th (see above), or available in Paper II. 

Throughout this discussion we classify different types of neutron-capture stars according to \citet{beers:05} as summarized 
in Table~\ref{tab:class}. In addition, we adopt the definition of s-II stars from Paper II. Compared with the s subclass, 
s-II stars actually are those with both strong r- and s- process enhancement ($\mathrm{[Ba/Fe]}>1.5$, $\mathrm{[Eu/Fe]}>1.0$ 
and $\mathrm{[Ba/Eu]}>0.5$).  

We note, HE1221-1948 is one of the several C-enhanced stars ([C/Fe]=1.42) in our analysis, which has been analysed 
by \citet{lucatello:06} manually, giving a metallicity of $-2.60$ and [Eu/Fe]=2.1. Since in Paper II, it was mentioned 
that our method is not suitable for C-rich stars spectra, and actually this star was noticed in figure 5 of Paper II 
with $\Delta\mathrm{[Fe/H]}>1$ comparing the final [Fe/H] with the initial estimate, thus in this work we demote this 
star to an upper limit, although $m' > 1$ was obtained, which we subjected to an overestimated $\mathrm{S/N}$. HE1221-1948 
is the only carbon rich one ($\mathrm{[C/Fe]}>1$) among the most metal-poor stars ($\mathrm{[Fe/H]}<-3.0$) with detected 
thorium abundances in our sample.
%$\mathrm{[Eu/Fe]}=2.1$\,dex for HE1221-1948, which has $\mathrm{[Fe/H]}=-3.36$ and $\mathrm{[C/Fe]}=1.42$ from our analysis, indicating it is a CEMP-r %%star ($\mathrm{[C/Fe]}>+1.0$ and $\mathrm{[Eu/Fe]}>+1.0$, see table 2 of \citet{beers:05}). Additionally, they derived a higher metallicity of $-2.60$, %which we take as the more reliable value, since  . 
%%%%%%%%%%%%%%%%%%%%%%%%%%%%%%%%%%%%%%%%%%%%%%%%%%%%%%%%%%%%%%%%%%%%%%%%%%%%%%%%%%%%%%%%%%%%%%%%%%%%%%%%%%%%%%%%%%%%%%%%%%%%%%%%%%%%%%
\subsection{Thorium abundances distribution}
\begin{figure}[htpb]
\resizebox{\hsize}{!}{\includegraphics[bb = 10 50 560 600, clip = true]{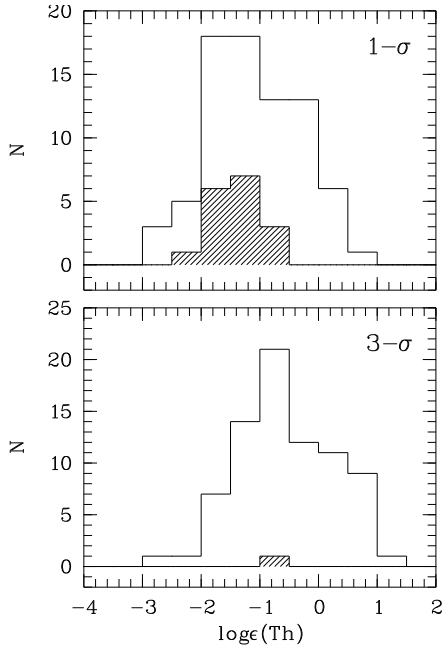}}
\caption{Histogram of derived thorium abundances. $1\sigma$ (upper panel) and $3\sigma$ (lower panel) detections 
are displayed. The shadow plots show the stars with detected abundances, while the blanks also include stars with 
upper limits.}    
\label{fig:thdist}
\end{figure}

In the sample of 253 stars, finally we obtain thorium abundances\footnote{Here we adopt the standard spectroscopic 
notations:\\
 $\mathrm{log}\epsilon(\mathrm{A})=\mathrm{log}_{10}(N_\mathrm{A}/N_\mathrm{H})+12$ for abundances;\\
 $\mathrm{[A/B]}=\mathrm{log}_{10}(N_\mathrm{A}/N_\mathrm{B})_\star-\mathrm{log}_{10}(N_\mathrm{A}/N_\mathrm{B})_\odot$ 
 for relative abundances;\\
log(A/B)=log$\epsilon$(A)-log$\epsilon$(B) for abundance ratios.} for 17 stars.  For another 60 stars, we obtained upper 
limits. In Fig.~\ref{fig:thdist}, we give the thorium abundances distribution of these stars. We were unable to derive 
the thorium abundance for the other 176 stars, 79 of which were set aside for various reasons, the most common being that 
the spectrum was too contaminated with blends to derive anything reliable given the resolution of the observed spectrum. 
We adopted $1\sigma$ detection level, and since our aim is to investigate the general trend of thorium abundances, this 
detection level is adequate to give a reliable distribution.

As shown in Fig.\ref{fig:thdist}, in the distribution of thorium abundances by the $1\sigma$ detection, majority of the 
stars lie in the range of $-2.0$ to $-1.0$, about $82\%$ of all the detections, and the distribution including $1\sigma$ 
upper limits extends from $-3.0$ to 1.0, having a peak at $-2.0$ to $-1.0$. The thorium abundance results in previous 
studies all fall in this range \citep[see e.g.,][]{roederer:09, honda:04}. In the lower panel of Fig.\ref{fig:thdist}, 
the only one detection above $3\sigma$ is owned by CS31082-001, which has $\mathrm{log}\epsilon\mathrm{(Th)}=-1.00$, 
and including those with $3\sigma$ upper limits the distribution spaning from $-3.0$ to 1.5, has a peak at the range 
of $-1.0$ to $-0.5$. Only the higher Th abundances side of this histogram can give useful information about the true 
shape of the cosmic distriubtion, since there is a natural detection bias towards larger abundances. The existence of 
a large scatter in thorium abundances among metal-poor stars as seen from this histogram can be explained by the poor 
mixing in the early universe. 

However, because the abundance of thorium is based on one single Th II line $4019$\,\AA~, it should be borne in mind 
the that Th abundance is susceptible to errors due the difficulties in modelling the blends. The line Th II $4019$\,\AA~ 
is blended with temperature sensitive spectral features, such as Co and Ni, thus, we expect some degree of the scatter 
is due to errors in the analysis, e.g., uncertain Teff, rather than real cosmic scatter. However, as shown in 
Fig.\ref{fig:thdist} and Fig.~\ref{fig:theufe} of this paper, the scatter significantly exceeds the estimated errors, 
and thus given the detection bias, it gives a lower limit to the range of the true cosmic distribution. 
  
%%%%%%%%%%%%%%%%%%%%%%%%%%%%%%%%%%%%%%%%%%%%%%%%%%%%%%%%%%%%%
\subsection{Comparation with previous studies}
\label{subsect:comparation}
\begin{figure}[htpb]
\resizebox{\hsize}{!}{\includegraphics[scale = 0.5, bb = 30 80 500 490, clip = true]{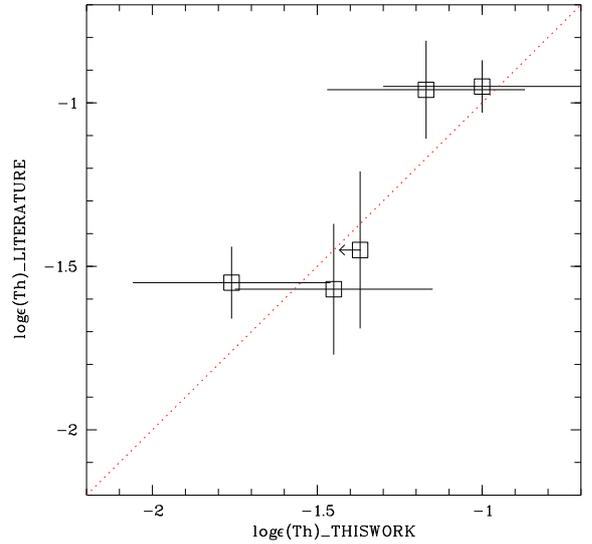}}
\caption{Thorium abundances comparison for the common stars between this work and previous works, which are the 
averaged value if 
there are more than one results. Error bars are given and upper limits are marked with arrows.}
\label{fig:thcom}
\end{figure}

As shown in Table \ref{tab:comp}, we obtained Th abundances for five stars which have been studied previously in detail 
with similar 
or better quality observational data by others, namely CS~22892$-$052, CS~29497$-$004, CS~31082$-$001, CS~29491$-$069 
and HE~2327$-$5642. Note that we failed to derive the thorium abundance for a well studied bright halo star HE~221170, 
since our modelling indicated Th II line in this star to be severely blended by Co lines; this star has been reported 
thorium enhanced compared with iron \citep{ivans:06,yushchenko:05}. The same for another strongly r-process enhanced 
star HE~1219$-$0312, in which the blendings exceed $80\%$ of the line fit, resulting a very uncertain abundance. 
HE~0338$-$3945 is an reported s-II carbon enhanced star ([Eu/Fe = 1.89, [Ba/Eu] = 0.52, and [C/Fe] = 2.07]), which 
experienced strongly both r- and s- process enhancements. For the same reason as for other carbon-enhanced stars, 
our method failed to get a reliable thorium abundance for this hot dwarf. For rest of the common stars, our thorium 
abundance results are all in agreement with literatures values within the uncertainty as shown in Fig.~\ref{fig:thcom}; 
such consistence gives us the confidence that our results are reliable for studying the overall thorium abundances 
distribution of metal-poor stars. 
\begin{table*}[htpb]
\centering
\caption{Comparison with literature results for seven common stars}
 \begin{tabular}{l c c c c c c c}
 \hline\hline  
   & $T_\mathrm{eff}$ & log$g$    & $\mathrm{[Fe/H]}$ & $V_\mathrm{mic}$   & $\mathrm{log}\epsilon(\mathrm{Th})$ & Notes \\
   &[K]               &[cm s$^{-2}$]&                &[km s$^{-1}$] &                                     &        \\
 \hline
 CS~22892$-$052& 4884& 1.81& $-2.95$& 1.67& $-1.76$        & This work\\
            & 4790& 1.6 & $-2.92$& 1.8 & $-1.42\pm0.15$ & \citet{honda:04}\\
            & 4800& 1.5 & $-3.1$ & 1.95& $-1.57\pm0.10$ & \citet{sneden:03}\\
            & 4710& 1.5 & $-3.2$ & 2.1 & $-1.60\pm0.07$ & \citet{sneden:00}\\
            & 4800& 1.5 & $-3.1$ & 1.95& $-1.60\pm0.13$ & \citet{roederer:09}\\
 \hline 
 CS~29497$-$004& 5013& 2.23& $-2.81$& 1.62& $-1.17$        & This work\\
            & 5090& 2.4 & $-2.81$& 1.6 & $-0.96\pm0.15$ & \citet{christlieb:04}\\  
 \hline
 CS~31082$-$001& 4922& 1.90& $-2.78$& 1.88& $-1.00$        & This work\\
            & 4825& 1.5 & $-2.9$ & 1.8 & $-0.98\pm0.05$ & \citet{hill:02}\\   
            &     &     & $-2.9$ &     & $-0.98\pm0.05$ & \citet{plez:04}\\
            & 4790& 1.8 & $-2.81$& 1.9 & $-0.92\pm0.10$ & \citet{honda:04}\\    
\hline
 CS~29491$-$069& 5103& 2.45& $-2.81$& 1.54& $<-1.37$       & This work\\
            & 5300& 2.8 & $-2.6$ & 1.6 & $-1.46\pm0.25$ & \citet{roederer:09}\\
            & 5300& 2.8 & $-2.6$ & 1.6 & $-1.43\pm0.22$ & \citet{hayek:09}\\
%\hline
% HE~0338$-$3945& 6162& 4.09& $-2.41$& 1.33& $<0.07$         & This work\\
%            & 6160& 4.13& $-2.42$& 1.13& $<0.23\pm0.19$ & \citet{jonsell:06}\\
\hline
 HE~2327$-$5642& 5048& 2.22& $-2.95$& 1.69& $-1.45$        & This work\\
            & 5050& 2.34& $-2.78$& 1.8 & $-1.67\pm0.20$ & \citet{mashonkina:10}\\  
\hline
 \end{tabular}
\label{tab:comp}
\end{table*}

\begin{figure}
\resizebox{\hsize}{!}{\includegraphics[scale = 0.5, bb = 30 100 550 660, clip = true]{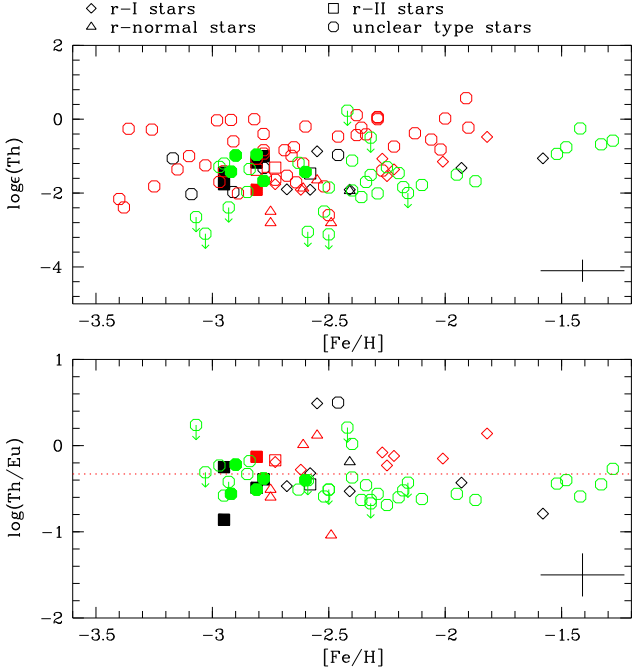}}
\caption{Plot of log$\epsilon$(Th) vs. [Fe/H] and log(Th/Eu) vs. [Fe/H]. Detected results from this work are in black, and the upper 
limits are in red. Results from other studies are in green. These filled symbols are the stars in common with previous works. See text
 for the definations of different subclasses. A dashed red line is plotted in the lower panel corresponding to the ''zero-age'' log(Th/Eu). Averaged error bars are given at the lower right corner.}  
\label{fig:theufe}
\end{figure}

In Fig.~\ref{fig:theufe}, we plot thorium abundances and the ratios of thorium over europium against increasing metallicity, 
and we also plot results collected from previous studies in green, including field halo 
stars \citep[e.g.,][]{cayrel:01,hill:02,frebel:07,roederer:09}, globular clusters M5 and M15 \citep{yong:08b,yong:08a},
 and a member of a nearby dwarf spheroidal galaxy \citep{aoki:07}. Filled symbols are used to mark the common stars listed
 in \ref{tab:comp}, while due to different metallicities adopted, they look not in pairs. In the lower panel a ''zero-age'' line 
 caculated based on the initial ratio $\mathrm{log}\epsilon\mathrm{(Th/Eu)}_\mathrm{{0}}=-0.33$ given by \citet{schatz:02} is shown, 
 and we will discuss more about age estimations in section \ref{subsec:ageestimation}. Stars with both detected and upper limits of 
 thorium abundances at the $\ge 1\sigma$ level are included. Europium and other element abundances in this work are from the analysis
 of Paper II. A summary of previous results on thorium abundances is given in online Table \ref{tab:otherworks}.

In the upper panel of Fig.~\ref{fig:theufe}, we see our sample provides detections which extend the range of derived thorium
 abundances to lower metallicities.  At $\mathrm{[Fe/H]}<-1.8$, a larger scatter can be seen, as expected according to the 
 scenario of poor mixing in the very early universe. No clear trend with increasing metallicity is seen, but the averaged 
 detected log$\epsilon$(Th) is $\sim0.4$ dex lower than those with metallicity above $-1.8$. 
Due to lack of derived Eu abundances, stars at the low metallicity end, e.g., HE~0353$-$6024 with the lowest metallicity of $-3.17$ 
in the upper panel were lost in the lower panel. The distribution of log(Th/Eu) ratios spans from $-0.86$ to 0.50, while considering the uncertainty, most of them are consistently close to the ''zero-age'' line, supporting the universal r-process pattern in the early universe, except for only two stars: r-I star HE~0105$-$6141 and HE~1332$-$0309 with no clear type yet, which have much higher
 log$\epsilon$(Th/Eu) of 0.49 dex and 0.50 dex, respectively. They might belong to the so-called ''actinide boost'' star group, 
 which we will discuss again in section \ref{subsec:ageestimation}.

\begin{table}
\centering
\caption{Definition of subclasses of metal-poor stars}
 \begin{tabular}{l l}
 \hline\hline  
Class & Constraints \\
\hline
 r-normal &  $0\leq\mathrm{[Eu/Fe]}<0.3$ and $\mathrm{[Ba/Eu]}<0$\\
\\
 r-I      &  $0.3\leq\mathrm{[Eu/Fe]}\leq1.0$ and $\mathrm{[Ba/Eu]}<0$\\
\\
 r-II     &  $\mathrm{[Eu/Fe]}>1.0$ and $\mathrm{[Ba/Eu]}<0$\\
\\
 s        &  $\mathrm{[Ba/Fe]}>1.0$ and $\mathrm{[Ba/Eu]}>0.5$\\
\\
 s-II     &  $\mathrm{[Eu/Fe]}>1.0$ and $\mathrm{[Ba/Eu]}>0.5$\\
\\
 r/s      &  $0.0<\mathrm{[Ba/Eu]}<0.5$    
\\
 \hline
 \end{tabular}
\label{tab:class}
\end{table}
%%%%%%%%%%%%%%%%%%%%%%%%%%%%%%%%%%%%%%%%%%%%%%%%%%%%%%%%%%%%%%%%

\subsection{Correlation with other neutron capture elements}
\label{subsect:correlation}
Correlations between abundances of thorium and other neutron-capture elements are used to explore the nature of nucleosynthesis, 
and impose contraints on astrophysical and nuclear models. Chemical abundance analysis suggest that for elements 
with $\mathrm{Z}=52\sim76$, the abundance pattern is consistent with the scaled solar system r-process abundance distribution.
Recent studies have found relatively low abundance levels for light elements ($\mathrm{Z}<50$) compared with heavy elements, 
which suggest two distinct types of r-process events, a main r-process for the elements above the second-peak elements, 
and a weak r-process for lighter neutron capture elements. The enhancement of actinides with respect to the rare earth elements
 were also noticed in some stars. 

\begin{figure*}
\begin{center}
\resizebox{\hsize}{!}{\includegraphics[scale = 0.5, bb = 0 130 630 350, clip = true]{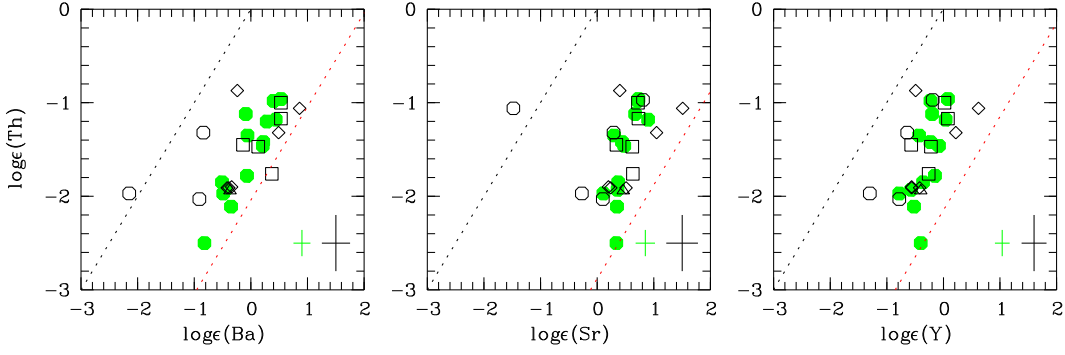}}
\end{center}
\caption{log$\epsilon$(Th) plotted against neutron capture elements: Ba, Sr and Y. A diagonal 1:1 relation (black), and the solar 
ratio (red) are plotted for comparision. Error bars of the results from this work and the literature are given in the lower right 
of each plot. The symbols are as same as in Fig.~\ref{fig:theufe}. HE~2219$-$0713 has higher Th than Ba, and HE~0353$-$6024 has higher 
Th than Sr.}
\label{fig:correlation1}
\end{figure*}

In this work, we also explored the correlations between thorium and other neutron-capture elements. In Fig.~\ref{fig:correlation1},
 we plot the ratios of Th abundances against Ba, Sr and Y. There are larger scatters among the results of Th correlated with Ba and Sr,  and usually the scatter comes from the unclear type stars, which indicates the existence of multiple nucleosynthesis prcesses. 
  Scatter of [Ba/Fe] and [Sr/Fe] were also reported by other studies \citep{mcwilliam:98,norris:01}.  All these elements follow
   similar abundance pattern as thorium as expected, and stars of the same enhancement type have quite consistent distribution,
    supporting that they might form from the materials experienced similar chemical enrichments. Elemental ratios of log(Th/Ba), log(Th/Sr) and log(Th/Y) for almost all the stars are enhanced compared with the ratios in solar system. Two stars: HE~2219$-$0713 and HE~0353$-$6024 are particular interesting. The former star has more Th than Ba, and the latter one has more Th than Sr. As shown in Fig. \ref{fig:spectra}, the Th II line is quite weak in both of the two stars, and even we can see some comparable ''noise'' or ''features'' unfitted, but since we failed to find the possible missing lines here, and our line list is believed to be well established, thus we suspect the S/N is overestimated for them, consequently leading to an overestimation of the detection significance. Better quality data for the two stars are needed to confirm this results.    

% This confirms that very different nucleosynthesis process happened in s-II and other r process stars. It has been suggested that
% the lighter elements might be synthesized in a ``weak'' r-process and the heavier elements might be produced in a more robust ``strong'' (or ``main'') r-process \citep{truran:02}. Therefore, do our results suggest that the ``weak'' r-process is common in all r-process stars, while the ``strong'' r-process might be more robust in some rare cases, e.g., in s-II stars? However, due to the incompleteness of abundances for the whole sample, the subsamples for each comparison are different, which we should keep in mind for before reaching any firm conclusion. 
 
\begin{figure*}
\begin{center}
\resizebox{\hsize}{!}{\includegraphics[scale = 0.5, bb = 0 210 625 420, clip = true]{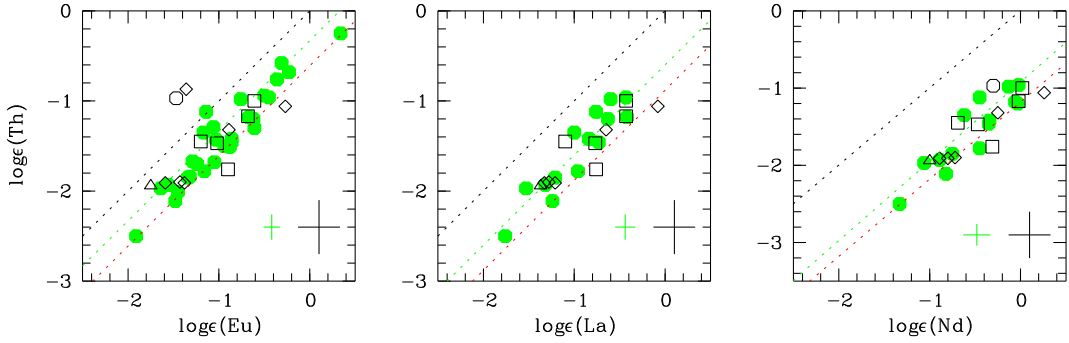}}
\end{center}
\caption{log$\epsilon$(Th) plotted against neutron capture elements: Eu, La and Nd. A diagonal 1:1 relation (black), the initial 
ratio (green)\citep{schatz:02} and the expected ratio assuming an age of 13 Gyr (red) are plotted for comparision. Error bars 
are also given in the lower right of each plot. The symbols are as same as in Fig.~\ref{fig:theufe}. The two stars that have much 
higher Th than Eu are HE~0105$-$6141 and HE~1332$-$0309.}
\label{fig:correlation2}
\end{figure*}

In Fig.~\ref{fig:correlation2}, the three chronometer ratios log(Th/Eu), log(Th/La) and log(Th/Nd) were plotted. Except the two stars with high 
enhancement of log(Th/Eu), all the other stars have consistent distribution with literature results, and the smaller scatter 
in the distributions, indicate the three age indicator will give consistent age estimations.\\
%\begin{figure}
%\resizebox{\hsize}{!}{\includegraphics[scale = 0.5]{thoriumf6.ps}}
%\caption{Absolute abundance ratios log(Th/Eu) plotted against light over heavy elments ratios. Symbols !%are as same as in Fig.~\ref{fig:theufe}.}  
%\label{fig:ratios}
%\end{figure}
%%%%%%%%%%%%%%%%%%%%%%%%%%%%%%%%%%%%%%%%%%%%%%%%%%%%%%%%
\subsection{Implications for age estimations and nuclear astrophysics}
\label{subsec:ageestimation}
The errors in Th abundance as well as the large scatter in the derived log(Th/Eu) ratios for metal-poor stars lead to significant 
uncertainties in age estimates. A $0.2$ dex error in log(Th/Eu) will induce an error of $9.3$ Gyr in age, which could be even more 
if considering the uncertainty in the predicted initial production ratios (PRs) \citep{frebel:07}. In our case, the error of 
log(Th/Eu) is estimated to be $>0.25$, which corresponds to an error of at least $>11.7$ Gyr in age determinations. 
In addition, it is known that, due to the so-called ``actinide boost'' phenomenon, the log(Th/Eu) chronometer pair fails to give
 meaningful estimates of age for some objects,  e.g., CS~30306$-$132, CS~31078$-$018, CS~31082$-$001 and HE~1219$-$0312,  
 all giving negative ages. These objects exhibit higher abundances of thorium and uranium with respect to the lanthanides,
  which is currently not understood. \citet{roederer:09} examined the Pb and Th abundances in 27 r-process only stars, and 
  suggest that, deviation from main r-process affect at most only the elements beyong the third r-process 
  peak elements, i.e., Pb, Th and U, while the reason for the low amount of Pb in CS~31082$-$001 is not clear yet. 
  Pb isotops and actinide elements measurements for more metal-poor stars are necessary for better understanding the 
  nucleosynthesis in such actinide enhanced stars, and to explore how common this ''actinide boost'' phenomenon may exist 
  in metal-poor stars.
%\begin{table*}
%\centering
%\caption{Objects with detected thorium abundances grouped into different subclasses of r-process enhanced metal-poor stars.}
% \begin{tabular}{c l}
% \hline\hline  
%Star & Teff \\
%\hline
%&&&&&&&&&&&\\
%\multicolumn{12}{c}{\underline{r-II stars}} \\
%&&&&&&&&&&&\\
%
%&&&&&&&&&&&\\
%\multicolumn{12}{c}{\underline{r-I stars}} \\
%&&&&&&&&&&&\\
%
%&&&&&&&&&&&\\
%\multicolumn{12}{c}{\underline{r-I stars}} \\
%&&&&&&&&&&&\\
%
% r-normal &  $0\leq\mathrm{[Eu/Fe]}<0.3$ and $\mathrm{[Ba/Eu]}<0$\\
%\\
% r-I      &  $0.3\leq\mathrm{[Eu/Fe]}\leq1.0$ and $\mathrm{[Ba/Eu]}<0$\\
%\\
% r-II     &  $\mathrm{[Eu/Fe]}>1.0$ and $\mathrm{[Ba/Eu]}<0$\\
%\\
%
% s-II     &  $\mathrm{[Eu/Fe]}>1.0$ and $\mathrm{[Ba/Eu]}>0.5$\\
%\\
% r/s      &  $0.0<\mathrm{[Ba/Eu]}<0.5$    
%\\
% \hline
% \end{tabular}
%\label{tab:class}
%\end{table*}

\begin{figure}
\begin{center}
\resizebox{\hsize}{!}{\includegraphics[bb = 45 15 555 720,clip = true]{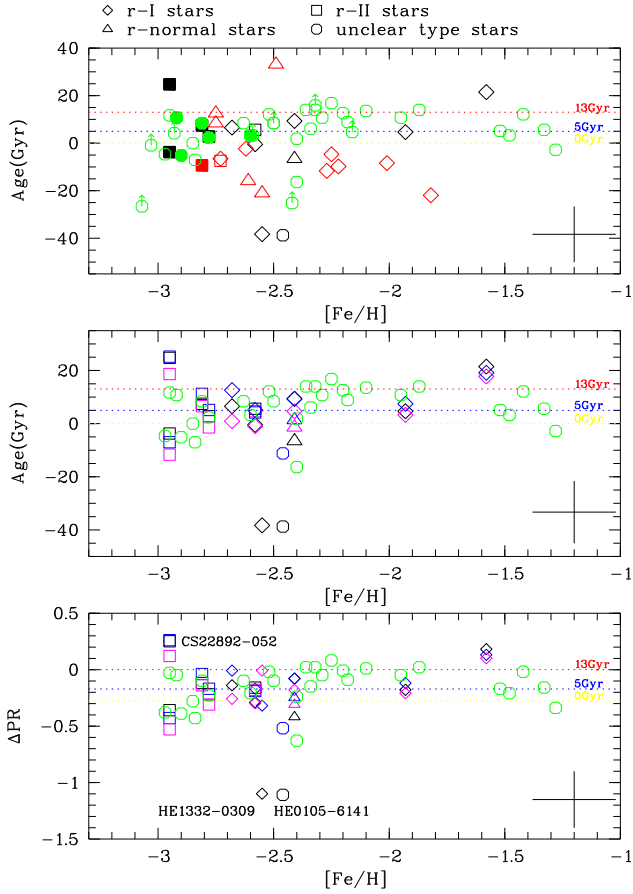}}
\end{center}
\caption{{\bf Upper Panel:} plot of ages derived from log(Th/Eu) against [Fe/H]; {\bf Middle Panel:} ages derived from different chronometers; 
{\bf Lower Panel:} plot of the differences between our expected initial production ratios and the theoretic predictions log(Th/Eu)$_0$ 
against [Fe/H] assuming age of 13Gyr for all the stars. Symbols and colours are as same as in Fig.~\ref{fig:theufe}, and in the 
middle panel, we use black, red and blue symbols for results derived from log(Th/Eu), log(Th/La) and log(Th/Nd) respectively.}  
\label{fig:age}
\end{figure}

Despite the large uncertainty in using log(Th/Eu) chronometer, it is meaningful to give the distribution of age estimations for 
so far the largest old population sample with avaliable log(Th/Eu). In Fig.~\ref{fig:age}, against [Fe/H], derived ages and the 
deviations of our expected initial production ratios (PRs), from the present theoretic initial PRs are plotted.
 We caculated the expected PRs assuming an age of $13$ Gyr for all the stars, since these stars are expected to be very old based 
 on their low metallicities. For age estimation we adopted $\mathrm{log(PR)}=-0.33$ 
 from \citet{schatz:02}, which is in the middle of other PRs \citep[][]{sneden:03,cowan:02,frebel:07}. Different PRs would only 
 change the absolute age determinations by smalle amounts, i.e., $\sim$2-4 Gyr \citep{frebel:07}. Stars (green) from previous 
 studies are also included for comparison. Ages from log(Th/Eu), together with that from  log(Th/La) and log(Th/Nd), if available, 
 based on the initial PRs of \citet{schatz:02}, are plotted in black, red and blue respectively in the middle panel.
  Lines corresponding to different ages or in the lower panel, the zero-deviation lines corresponding to different age chronometers  are plotted. 

It can be seen from the plot that, the ages derived from log(Th/Eu) for those metal-poor stars in our sample are in general 
consistent with literature values within the uncertainty, except for HE~0105$-$6141 and HE~1332$-$0309, and there is no apparent
 trend with increasing metallicity. It's seen that the derived ages distribute around 13 Gyr and scatter below 0 Gyr, 
 with several ''negative age'' stars as previously reported from other works, and also in this work. In upper panel of Fig. \ref{fig:theufe}, HE~0105$-$6141 and 
 HE~1332$-$0309 are in the higher end of continuous log$\epsilon$(Th) distribution, but the ratio of log(Th/Eu) reach as high 
 as $\sim$~0.5 dex, $\sim0.5$ dex higher than other stars. Thus, if considering the limitations of our data set, only these 
 two stars are very likely to be ''actinide boosted''.

\begin{figure*}
%\resizebox{\hsize}{!}{\includegraphics{fff6.eps}
\resizebox{\hsize}{!}{\includegraphics[bb = 20 350 550 1055,clip = true]{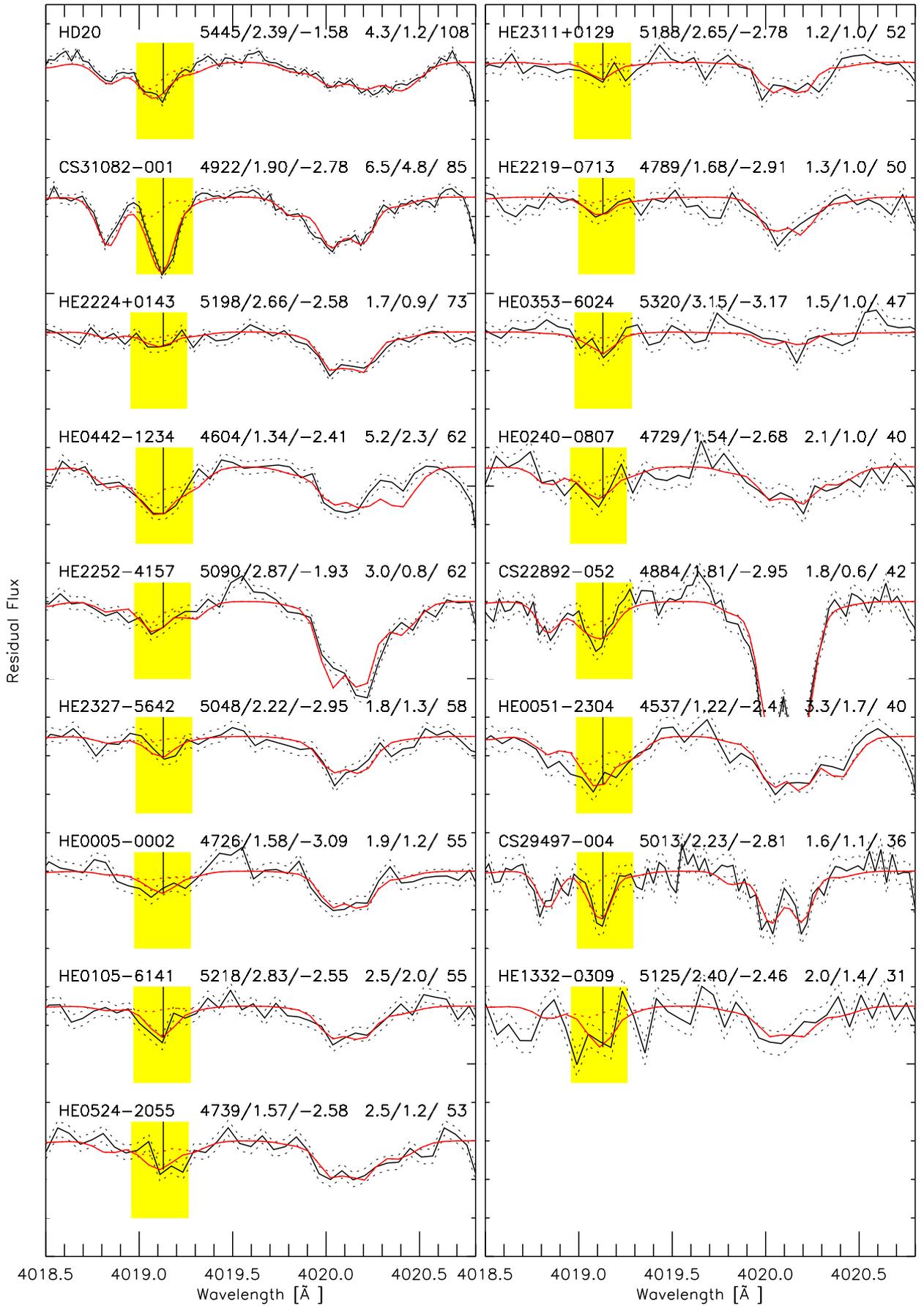}}
%\resizebox{\hsize}{!}{\includegraphics[bb = 5 345 590 1100,clip = true]{3f6.ps}}
\caption{Th II line profile fits for stars with detected thorium abundances above $1\sigma$. Fitting regions are highlighted in bright yellow. 
{\bf Black Solid:} observed spectrum; {\bf Black Dotted:} observed spectrum including noise; {\bf Red Solid:} 
 best fit synthetic spectrum; {\bf Red Dotted:} best synthetic spectrum after removing the \ion{Th}{II} line.}
\label{fig:spectra}
\end{figure*}

\begin{figure*}
\begin{center}
%!\resizebox{\hsize}{!}{\includegraphics[scale = 0.5, bb = 35 270 545 490,clip = true]{fff7ps.ps}}
\resizebox{\hsize}{!}{\includegraphics[scale = 0.5, bb = 35 240 545 580,clip = true]{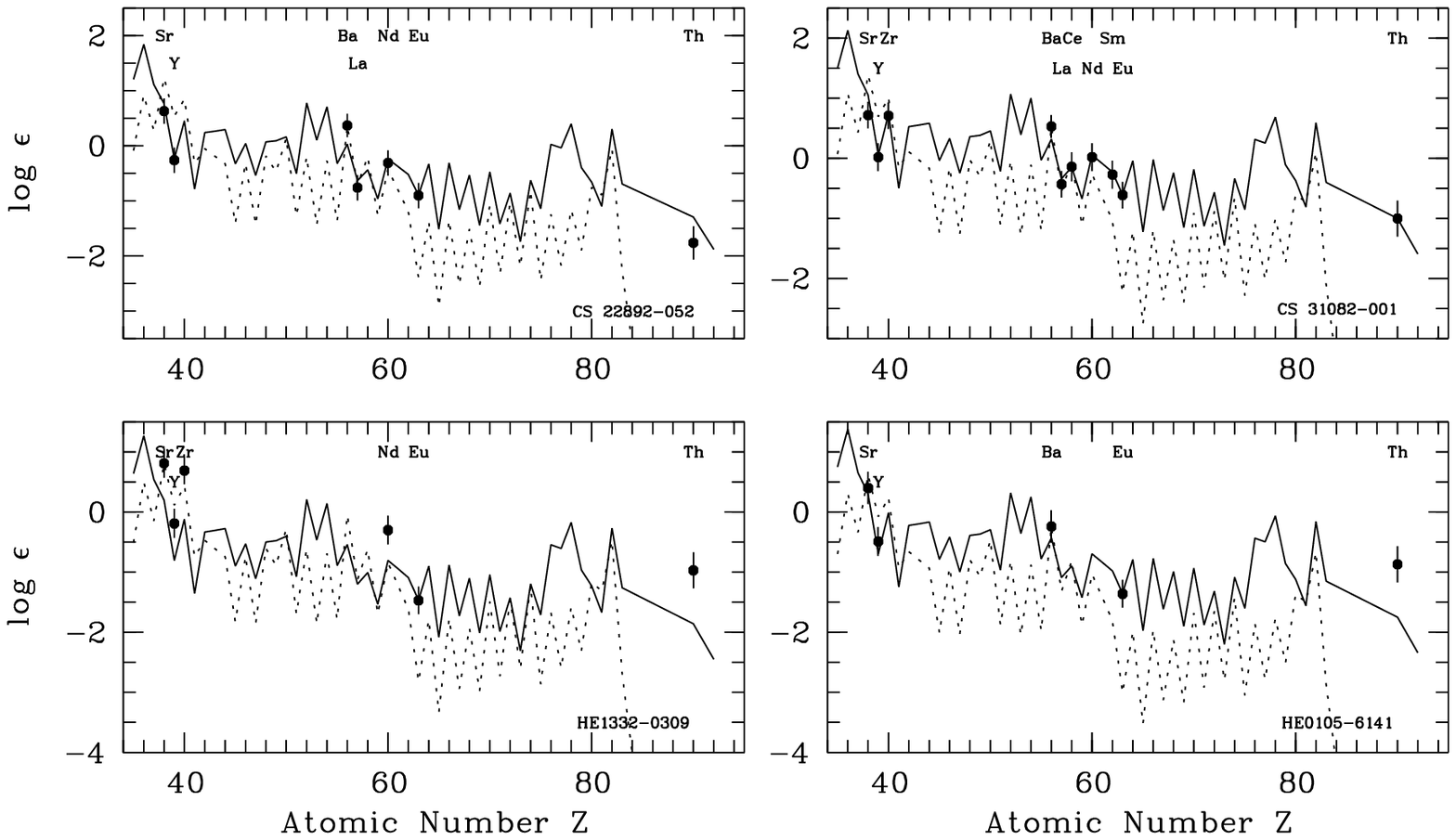}}
\end{center}
%\caption{The abundance pattern of CS~22892$-$052 (black); CS~31082$-$001 (red); HE~0131$-$3953 (green); HE~1247$-$2114 (blue); HE~1132$-$0204 (yellow) and HE~0105$-$6141 (magenta), compared with the solar r-process pattern (solid line) scaled to the Eu abundance of CS~22892$-$052, and the solar s-process pattern (dashed line) scaled to the Ba abundance of the same star. The r- and s- fractions are from \citet{arlandini:99}, except for Th and U which are from \citet{burris:00}.}  
\caption{Abundance pattern of CS~22892$-$052, CS~31082$-$001, HE~1332$-$0309 and HE~0105$-$6141, compared with the solar
 r-process pattern (solid line) scaled to the Eu abundance and the solar s-process pattern (dotted line) scaled to the Ba or Sr 
 abundance of each star. The r- and s- fractions are from \citet{arlandini:99}, except for Th and U which are from \citet{burris:00}.}  
\label{fig:solarpattern}
\end{figure*}

In Fig. \ref{fig:spectra}, the \ion{Th}{II} line profile fits for stars above $1\sigma$ detection are displayed. The observed spectra 
are in solid black lines, and the noise on the spectra were overplotted in dotted black lines. A range of 3 \AA~ fit region is
 highlighted in bright yellow. The red solid lines are the best fit synthetic profile and the red dotted lines are the
  profile fits with only blendings, from which it's easily seen how severe the \ion{Th}{II} lines are blended. Star name, parameters, 
  detection confidence $m$ and $m'$, and S/N are shown in the top of each plot. 
 HE~1332$-$0309 has a low S/N of 31, and the observed line profile could't be reproduced well with present line list, which may 
 due to low S/N or unresolved blendings in the Th II line region. HE~0105$-$6141 with higher S/N of 55, however also suffers from
  noise. We notice the blendings around Th II for the two stars are very light, which may also indicate that, some missing blending
   components might lead to an overestimation of Th abundance. Thus, better understanding of the blendings and higher quality
    data of these objects need to be obtained in the future to verify this preliminary result. \\
For the common star CS~31082$-$001, we derived $\mathrm{log(Th/Eu)}=-0.39$, $\mathrm{log(Th/La)}=-0.57$, $\mathrm{log(Th/Nd)}=-1.02$, 
which is consistent with $-0.22$, $-0.60$, and $-0.91$ from \citet{schatz:02} within the uncertainty. Although we got a very consistent Th abundance with \citet{hill:02}, we adopted a 0.15 dex higher Eu abundance from Paper II, which leads the offset 
 in log(Th/Eu).  We adopted the zero-age  log(Th/X)$_0$ (X is the stable r-process elements) ratios from \citet{schatz:02}, 
 where they used solar abundances X$_{\sun}$  instead of the r-process model predictions X$_0$, therefore, the discrepancies 
 in the estimated ages from different log(Th/X) reflect deviations of the observed stellar abundances from a solar abundance pattern.  
As shown in the middle panel of Fig. \ref{fig:age}, the deviations are tiny and similar for most stars, for which two or three 
chronometers are available, and log(Th/La) always gives the youngest ages, indicating compared with Eu and Nd, observed La abundances
 always have larger deviations from solar La abundance. For HE~0105$-$6141, only log(Th/Eu) and log(Th/Nd) are avaliable, the latter one gives a much older age, which may suggest the high log(Th/Eu) in this stars is due to an abnormal low Eu abundance, while so far no other studies give Eu abundance for this star.\\
From the bottom panel of Fig. \ref{fig:age}, it shows if assuming a consistent age of 13 Gyr for these metal-poor stars, their 
intial production ratios for most of the stars seems needed to be increased by a factor up to 4, and even more if the 
two ''actinide boost'' stars are confirmed. \\
In order to explore more about these two unique stars, in Fig. \ref{fig:solarpattern}, we plot their abundance patterns, 
compared with solar r- and s-process patterns scaled to Eu and Ba (or Sr) respectively. For HE~1332$-$0309, because no Ba
abundance has been obtained, we scaled the solar s-process pattern to Sr. As comparison, a thorium normal r-II star CS22892$-$052, 
a previous reported ``actinide boost'' r-II star CS~31082$-$001 are also plotted. It can be seen from the plot, except for Th, the 
avaliable neutron-capture abundances of CS22892$-$052 follow the Solar pattern quite well, and the same is for CS~31082$-$001,
 but for both stars, we underestimated their log(Th/Eu) ratios. The situation is different for HE~0105$-$6141, for which the Th abundance was highly enhanced, but the other avaliable neutron-capture elements follow the Solar r-process pattern. For HE~1332$-$0309, Barium abundance is not avaliable. Beside Th, Nd, Sr, Zr, and Y are also enhanced to some degree compared with solar r-process
  pattern.  \\
%%%%%%%%%%%%%%%%%%%%%%%%%%%%%%%%%%%%%%%%%%%%%%%%%%%%%%%%%%%%%%%%%%%%%%%%%%%%%%%%%%%%%%%%%%%%%%%%%%%%%%%%%%%%%%%%%%%%%%%%%%%%%%%
\section{Conclusion}
\label{sect:conclusion}
Using the ``snapshot'' spectra from HERES survey, we derived the thorium abundances of 77 metal-poor stars, 17 of which have 
detected Th abundance, while for the rest, only upper limits are available. Thorium abundances cover a wide range of about 4.0 dex, 
and a scatter exists in the distribution of log(Th/Eu) ratios, supporting previous studies suggesting that the r-process is not universal. For the five common stars, our results are in good agreement with previous studies, which gives us the confidence to present 
 a reliable Th distribution and to discuss r-process pattern for such a large sample of metal-poor stars.  
 
With avaliabe abundances, we explored the correlation between Th and other r-process elements. We confirmed the relatively large
 scatter in log(Th/Ba) and log(Th/Sr) distributions, and found that a better consistence exists within r-process stars, which confirms that these stars formed from the gas experienced similar nucleosynthesis.
  
Using log(Th/Eu) as the chronometer, we derived the ages for r-process metal-poor stars, for which both Th and Eu are available.
Two stars might be ''actinide boosted'', considering an error to 11.7 Gyr in our age estimation. These stars might have experienced 
very different chemical enrichments during their formation and evolution. At present the only possible explanation would be that, 
the r-process elements in these stars were implanted long after the formation of the star, still holding the principle of a universal 
r-process pattern. It suggests that ``actinide boost'' might not be a rare
   phenomenon, thus questioning the reliability in using log(Th/$\mathrm{X}_{\mathrm{stable}}$) for cosmochronometry to derive stellar ages.
 For this kind of stars, U abundances are needed to confirm the ''actinide boost'' feature and log(Th/U) can be used to 
give more accurate age estimations.

However, it should be cautioned again that, $1\sigma$ detection significance was adopted in this work, due to the weakness of Th II line and
 the limited quality of our present data, which may lead to significant uncertainty in the
 thorium abundance. Thus, the detection of Th in stars without enhancement of other n-capture elements and the ''actinide boost'' 
 phenomenon in some stars are not confirmed conclusion. Better quality spectra are needed for higher precision Th and other 
 r-process heavy elements abundance determinations.
 Thus here we would not like to give any conclusion, but rather to draw attention on those very interesting objects from this work,
   i.e., HE~2219$-$0713, HE~0305$-$6024, HE~0105$-$6141 and HE~1332$-$0309. 
     Actually we ourselves have already submitted a proposal for the observation time using ESO VLT/UVES for obtaining better quality 
     data of these objects, 
    if some of them can be confirmed, it will be a very important discovery. 

\begin{table*}
\centering
\caption{Stars with detected Th abundances classified into different subclasses of metal-poor stars.}
\label{tab:thdetect}
\begin{tabular}{lcrrrrrr}
\hline\hline
        Name & $\mathrm{T}_\mathrm{eff}$ &  log$g$        & [Fe/H] & log$\epsilon$(Th) & [Th/Fe] & log(Th/Eu)& "Age" \\
             & [K]                       & [cm s$^{-2}$]  &        &                   &         &           & [Gyr] \\
\hline
&&&&&&&\\   
\multicolumn{8}{c}{\underline{r-I stars}} \\ 
&&&&&&&\\      
HD~20          & 5445 &  2.39 & $-1.58$ & $-1.06$  & 0.43 & $-0.79$ & 23.8 \\
HE~0105$-$6141 & 5218 &  2.83 & $-2.55$ & $-0.87$  & 1.59 & 0.49    & $-38$ \\ 
HE~0240$-$0807 & 4729 &  1.54 & $-2.68$ & $-1.90$  & 0.69 & $-0.47$ & $6.5$ \\
HE~0442$-$1234 & 4604 &  1.34 & $-2.41$ & $-1.91$  & 0.41 & $-0.53$ & 9.3 \\
HE~0524$-$2055 & 4739 &  1.57 & $-2.58$ & $-1.91$  & 0.58 & $-0.32$ & $-0.5$   \\
HE~2252$-$4157 & 5090 &  2.87 & $-1.93$ & $-1.32$  & 0.52 & $-0.43$ &  4.7  \\
%HE~0315$+$0000 & 5013 &  2.11 & $-2.73$ & $-1.61$  & 1.03    & $-0.04$ & $-13.5$ \\
%HE~0328$-$1047 & 5301 &  3.03 & $-2.25$ & $-1.12$  & 1.04    &  0.19   & $-24.3$ \\
%HE~0501$-$5644 & 5033 &  2.30 & $-2.41$ & $-1.42$  & 0.90    &  0.17   & $-23.3$\\
%HE~1311$-$1412 & 4796 &  1.50 & $-2.91$ & $-1.80$  & 1.02    & 0.04    & $-17.3$ \\ 
%HE~2134$+$0001 & 5257 &  3.00 & $-2.22$ & $-1.12$  & 1.01    & 0.12    & $-21$ \\ 
&&&&&&&\\   
\multicolumn{8}{c}{\underline{r-II stars}} \\ 
&&&&&&&\\ 
CS~22892$-$052 & 4884 &  1.81 & $-2.95$ & $-1.76$  & 1.10    & $-0.86$ &  24.7 \\
CS~29497$-$004 & 5013 &  2.23 & $-2.81$ & $-1.17$  & 1.55    & $-0.49$ &  7.5 \\
CS~31082$-$001 & 4922 &  1.90 & $-2.78$ & $-1.00$  & 1.69    & $-0.39$ &  2.8 \\
HE~2224$+$0143 & 5198 &  2.66 & $-2.58$ & $-1.47$  & 1.02    & $-0.45$ &  5.6 \\      
HE~2327$-$5642 & 5048 &  2.22 & $-2.95$ & $-1.45$  & 1.41    & $-0.25$ &  $-3.7$ \\
&&&&&&&\\   
\multicolumn{8}{c}{\underline{r-normal stars}} \\ 
&&&&&&&\\ 
HE~0051$-$2304 & 4537 & 1.22 & $-2.41$ & $-1.94$ & 0.38 & $-0.19$ & $-6.5$ \\
%HE~1132$+$0204 & 5046 &  2.25 & $-2.55$ & $-1.20$  & 1.26    &  0.57   &  $-42$ \\
%HE~1225$+$0155 & 4842 &  1.80 & $-2.75$ & $-2.15$  & 0.51    & $-0.16$ &  $-7.9$ \\
%HE~1247$-$2114 & 5012 &  2.08 & $-2.61$ & $-1.43$  & 1.09    & 0.45    &  $-36.4$ \\ 
&&&&&&&\\   
%\multicolumn{8}{c}{\underline{s-II stars}} \\ 
%&&&&&&&\\ 
%HE~0131$-$3953 & 5928 &  3.83 & $-2.71$ &  0.15    & 2.77    & 0.70    & - \\ 
%&&&&&&&\\   
\multicolumn{8}{c}{\underline{unclassified stars}} \\ 
&&&&&&&\\ 
HE~0005$-$0002 & 4726 &  1.58 & $-3.09$ & $-2.03$  & 0.97    & -  & - \\
%HE~0029$-$1839 & 5010 &  2.19 & $-2.50$ & $-1.80$  & 0.61    & -       & - \\ 
%HE~0143$-$1135 & 5629 &  4.53 & $-2.13$ & $-0.25$  & 1.79    & -       & - \\
%HE~0331$-$4939 & 5206 &  2.60 & $-2.90$ & $-1.16$  & 1.65    & -       & - \\
%HE~0333$-$4001 & 5892 &  4.31 & $-2.64$ & $-0.05$  & 2.50    & -       & - \\
HE~0353$-$6024 & 5320 &  3.15 & $-3.17$ & $-1.06$  & 2.02    & -  & - \\
%HE~0441$-$4343 & 5629 &  3.59 & $-2.52$ & $-0.53$ & 1.90     & -       & - \\
%HE~0520$-$1748 & 5272 &  3.06 & $-2.52$ & $-1.10$  & 1.33    & -       & - \\
%HE~1243$-$1425 & 5507 &  3.19 & $-2.67$ & $-0.65$ & 1.93     & -       & - \\
%HE~1246$-$1344 & 4853 &  1.65 & $-3.40$ & $-2.03$  & 1.28    & -       & - \\
%HE~2156$-$3130 & 4692 &  1.28 & $-3.13$ & $-1.86$  &  1.18   & -       & - \\
HE~1332$-$0309 & 5125 & 2.40 & $-2.46$ & $-0.97$ & 1.40    & 0.5      & $-38.7$ \\
HE~2219$-$0713 & 4789 &  1.68 & $-2.91$ & $-1.97$  & 0.85    & -  & - \\
HE~2311$+$0129 & 5188 &  2.65 & $-2.78$ & $-1.32$  & 1.37    & -  & - \\
%HE~2345$-$1919 & 5617 &  4.46 & $-2.46$ & $-0.19$  & 2.18    & -       & - \\
\hline
\end{tabular}
\end{table*}
\begin{acknowledgements} 
The authors would like to give thanks to Dr. Paul Barklem for improving and running the code, for modifying and commenting
 on the paper draft, especially for his contribution to writing Sect. \ref{sect:anay}, as well as for his enthusiasm on this work.      
J.R. and N.C. acknowledge financial support by the Research Links program
of the Swedish Research Council, the Global Networks program of
Universit\"at Heidelberg, and by Deutsche Forschungsgemeinschaft through
 grant CH~214/5-1 and Sonderforschungsbereich SFB 881 “The Milky Way
 System” (subprojects A4 and A5).   
J.R. and G.Z. acknowledge the support by NSFC grant No. 10821061.
\end{acknowledgements}

\bibliographystyle{aa}
\bibliography{atomdata,cosmology,sdss,techniques}
%%%%%%%%%%%%%%%%%%%%%%%%%%%%%%%%%%%%%%%%%%%%%%%%%%%%%%%%%%%%%%%%%%%%%%%%%%%%%%%%%%%%%%%%%%%%%%%%%%%%%%%%%%%%%%%%%%%%%%%%

\Online

%\appendix

%\longtab{1}{
%\begin{landscape}
%\LTcapwidth=\linewidth
\LTcapleft=0pt
\setlength{\tabcolsep}{1mm}
\begin{longtable}{lrrrrrrrrrrcc}
%\begin{scriptsize}
\caption{Summary of abundance results for stars with Th abundances from this work. Except for Th, other element abundances and the stellar parameters are from Paper II. The errors are discussed in the text. $m$ and $m'$ are the detection 
levels for \ion{Th}{II} 4019.129 line profile fit with and without blendings.}\label{tab:result}\\
\hline\hline
Name&$T_\mathrm{eff}$&log$g$&[Fe/H]&$V_\mathrm{mic}$&log$\epsilon$(Th)&[Th/Fe]&[Eu/Fe]&[Ba/Eu]&log(Th/Eu)&[C/Fe]& $m(\sigma$)& $m'(\sigma$)\\ 
    & [K]            &[cm s$^{-2}$]& &[$\mathrm{kms^{-1}}$]&            &       &       &       &          &      &          & \\ 
\hline  
\endfirsthead
\caption{Continued.} \\  
\hline\hline
Name&$T_\mathrm{eff}$&log$g$&[Fe/H]&$V_\mathrm{mic}$&log$\epsilon$(Th)&[Th/Fe]&[Eu/Fe]&[Ba/Eu]&log(Th/Eu)&[C/Fe]& m($\sigma$)&m'($\sigma$)\\ 
    & [K]            &[cm s$^{-2}$]& &[$\mathrm{kms^{-1}}$]&            &       &       &       &          &      &          & \\ 
\hline
\endhead
\hline
\endfoot
\smallskip               
CS~22175$-$007 & 5108 &  2.46 & $-2.81$ & 1.67 & $<-1.07$ & $<1.65$ & -    & -       & -       & 0.19    & 0.39 & 0.15 \\               
%CS~22186$-$023 & 5066 &  2.19 & $-2.72$ & 1.58 & $<-1.53$ & $<0.23$ & -    & -       & -       & 0.30    & 0.35 \\               
%CS~22186$-$025 & 4985 &  1.70 & $-2.87$ & 2.14 & $<-1.35$ & $<0.29$ & -    & -       & -       & $-0.68$ & 0.18 \\               
CS~22886$-$042 & 4881 &  1.85 & $-2.68$ & 1.84 & $<-1.53$ & $<1.06$ & -    & -       & -       & 0.01    & 0.67 & 0.48 \\               
CS~22892$-$052 & 4884 &  1.81 & $-2.95$ & 1.67 & $-1.76$  & 1.10    & 1.54 & $-0.35$ & $-0.86$ & 1.00    & 1.81 & 0.61 \\        
CS~22945$-$028 & 5126 &  2.55 & $-2.66$ & 1.53 & $<-0.99$ & $<1.58$ &  -   & -       & -       & 0.21    & 0.49 & 0.33 \\              
%CS~22957$-$013 & 4904 &  1.96 & $-2.64$ & 1.79 & -        & -       & -    & -       & -       & 0.10    & \\              
%CS~22958$-$083 & 5101 &  2.40 & $-2.79$ & 1.50 & -        & -       & -    & -       & -       & 0.64    & \\               
%CS~22960$-$010 & 5737 &  4.85 & $-2.65$ & 1.53 & $<0.20$  & $<1.36$ & -    & -       & -       & 0.82    & 0.11 \\              
CS~29491$-$069 & 5103 &  2.45 & $-2.81$ & 1.54 & $<-1.37$ & $<1.35$ & 1.06 & $-0.72$ & $<-0.13$& 0.18    & 0.45 & 0.20 \\       
%CS~29491$-$109 & 4736 &  1.50 & $-2.90$ & 1.82 & -        & -       & -    & -       & -       & $-0.19$ &   \\           
CS~29497$-$004 & 5013 &  2.23 & $-2.81$ & 1.62 & $-1.17$  & 1.55    & 1.62 & $-0.41$ & $-0.49$ & 0.22    & 1.63 & 1.12 \\       
CS~29510$-$058 & 5108 &  2.32 & $-2.61$ & 1.62 & $<-1.20$ & $<1.32$ & -    & -       & -       & 0.40    &0.24  & 0.08  \\               
%CS~30308$-$035 & 4806 &  1.78 & $-3.35$ & 1.64 & -        & -       & -    & -       & -       & 0.04    &  \\              
%CS~30315$-$001 & 4565 &  1.14 & $-2.98$ & 1.99 & $<-2.09$ & $<0.46$ & -    & -       & -       & $-0.50$ & 0.63 \\               
%CS~30315$-$029 & 4541 &  1.07 & $-3.33$ & 2.06 & -        & -       & 0.72 & $-0.30$ & -       & $-0.43$ & \\              
%CS~30337$-$097 & 4865 &  1.81 & $-2.73$ & 1.88 & -        & -       & -    & -       & -       & $-0.02$ & \\              
%CS~30339$-$041 & 5478 &  2.10 & $-2.20$ & 2.23 & $<-0.93$ & $<0.29$ & -    & -       & -       & $-0.41$ & 0.31 \\               
%CS~31060$-$047 & 4749 &  1.55 & $-2.72$ & 2.00 & -        & -       & -    & -       & -       & $-0.26$ &\\               
%CS~31062$-$041 & 4915 &  1.83 & $-2.67$ & 1.84 & $<-1.61$ & $<0.22$ & -    & -       & -       & 0.49    & 0.48 \\               
CS~31082$-$001 & 4922 &  1.90 & $-2.78$ & 1.88 & $-1.00$  & 1.69    & 1.66 & $-0.48$ & $-0.39$ & 0.22    & 6.50 & 4.81 \\
HD~20          & 5445 &  2.39 & $-1.58$ & 2.30 & $-1.06$  &  0.43   & 0.80 & $-0.49$ & $-0.79$ & $-0.34$ & 4.27 & 1.35 \\
%HE~2335$-$5958B & 5765 &  3.24 & $-2.33$ & 1.63 & $<-0.48$ & $<-0.18$ &  -   & -       & -       & 0.08    & 0.12  \\             
HE~0005$-$0002 & 4726 &  1.58 & $-3.09$ & 1.82 & $-2.03$  & 0.97    & -    & -       & -       & 0.17    & 1.90 & 1.19 \\              
HE~0023$-$4825 & 5816 &  3.63 & $-2.06$ & 1.45 & $<-0.55$ & $<1.42$ & -    & -       & -       & 0.31    & 0.61 & 0.16 \\              
HE~0029$-$1839 & 5010 &  2.19 & $-2.50$ & 1.67 & $<-2.60$  & $<0.19$& -    & -       & -       & 0.31    & 1.13 & 0.28 \\              
%HE~0037$-$2657 & 4982 &  2.21 & $-3.22$ & 1.80 & $<-1.82$ & $0.45$  & -    & -       & -       & 0.31    & 0.42 \\              
HE~0039$-$4154 & 4735 &  1.55 & $-3.38$ & 2.01 & $<-2.39$ & $<0.9 $ & -    & -       & -       & $-0.14$ & 0.84 & 0.40 \\             
HE~0043$-$2845 & 5517 &  4.42 & $-2.91$ & 1.17 & $<-0.60$ & $<2.22$ & -    & -       & -       & 0.19    & 0.21 & 0.14 \\              
%HE~0044$-$2459 & 5242 &  2.92 & $-3.28$ & 1.55 & -        & -       & -    & -       & -       & 0.45    & \\             
%HE~0044$-$4023 & 5694 &  3.26 & $-2.56$ & 1.49 & -        & -       & -    & -       & -       & 0.40    & \\             
%HE~0045$-$2430 & 5377 &  3.25 & $-1.77$ & 1.67 & -        & 0.49    & 0.33 & $-0.67$ & $-0.26$ & $-0.09$ &  \\              
%HE~0049$-$5700 & 5952 &  4.08 & $-2.41$ & 1.63 & $<-0.11$ & $<1.89$ &  -   & -       & -       & 0.39    & 0.54 \\ 
HE~0051$-$2304  & 4537 &  1.22 & $-2.41$ & 1.97 & $-1.94$  & 0.38   & 0.17 & $-0.24$  & $-0.19$ & $-0.64$ & 3.30 & 1.71 \\         
HE~0054$-$0657 & 5908 &  4.40 & $-2.00$ & 1.56 & $<0.02$ & $<1.93$ & -    & -        & -       & 0.29    & 0.43 & 0.10 \\              
%HE~0057$-$4541 & 5083 &  2.55 & $-2.32$ & 1.67 & -        & -       & 0.62 & $-0.73$ & -       & 0.17    &\\               
%HE~0104$-$4007 & 5154 &  2.64 & $-3.30$ & 1.72 & $<-1.33$ & $<1.62$ & -    & -       & -       & 0.50    & 0.63 \\               
HE~0105$-$6141 & 5218 &  2.83 & $-2.55$ & 1.66 & $-0.87$  & 1.59    & 0.68 & $-0.51$ &  0.49   & 0.20    & 2.48 & 2.10 \\              
%HE~0109$-$0742 & 5315 &  1.85 & $-2.53$ & 2.19 & $<-1.39$ & $0.29$  & -    & -       & -       & $-0.14$ & 0.38 \\              
HE~0109$-$3711 & 6156 &  3.91 & $-1.91$ & 1.60 & $<0.57$ & $<2.39$ & -    & -       & -       & 0.31    & 0.28 & 0.14 \\              
HE~0121$-$2826 & 4955 &  1.99 & $-2.97$ & 1.68 & $<-1.70$ & $<1.18$ & -    & -       & -       & 0.54    & 0.32 & 0.12 \\             
%HE~0131$-$2740 & 5351 &  3.03 & $-3.08$ & 1.62 & $<-0.54$ & $<1.52$ & -    & -       & -       & 0.35    & 0.17 \\               
%HE~0131$-$3953 & 5928 &  3.83 & $-2.71$ & 1.06 &  0.15    & 2.77    & 1.62 &  0.58   & 0.70    & 2.45    & 2.50\\              
HE~0143$-$1135 & 5629 &  4.53 & $-2.13$ & 1.42 & $<-0.38$  & $<1.66$& -    & -       & -       & 0.23    & 0.72 & 0.25 \\              
%HE~0143$-$4108 & 5121 &  2.41 & $-2.62$ & 1.51 & -        & -       & -    & -       & -       & 0.16    &\\              
%HE~0231$-$4016 & 5972 &  3.59 & $-2.08$ & 1.67 & $<-0.23$ & $<1.76$ & -    & -       & -       & 1.36    & 1.03\\              
HE~0240$-$0807 & 4729 &  1.54 & $-2.68$ & 1.96 & $-1.90$  & 0.69    & 0.73 & $-0.52$ & $-0.47$ & $-0.35$ & 2.10 & 1.05 \\              
%HE~0240$-$6105 & 4720 &  1.26 & $-3.23$ & 2.15 & -        & -       & -    & -       & -       & $-0.25$ &\\              
%HE~0243$-$0753 & 5066 &  2.29 & $-2.49$ & 1.77 & -        & -       & -    & -       & -       & 0.29    &\\              
%HE~0243$-$5238 & 5085 &  2.35 & $-3.04$ & 1.64 & $<-1.87$ & $<0.99$ & -    & -       & -       & 0.40    & 0.93 \\              
%HE~0244$-$4111 & 5624 &  3.39 & $-2.56$ & 1.38 & $<-0.88$ & $<1.56$ & -    & -       & -       & 0.25    & 0.96 \\              
%HE~0248$+$0039 & 5199 &  2.57 & $-2.53$ & 1.75 & -        & -       & -    & -       & -       & 0.09    &\\              
%HE~0256$-$1109 & 5891 &  4.04 & $-2.73$ & 1.69 & -        & -       & -    & -       & -       & 0.67    &\\              
HE~0300$-$0751 & 5280 &  2.97 & $-2.27$ & 1.61 & $<-1.07$ & $<1.11$ & 0.77 & $-0.75$ & $<-0.08$& 0.10    & 0.98 & 0.61 \\              
%HE~0305$-$4520 & 4817 &  1.56 & $-2.91$ & 2.02 & -        & -       & -    & -       & -       & 0.33    &\\              
HE~0315$+$0000 & 5013 &  2.11 & $-2.73$ & 1.72 & $<-1.76 $ & $<0.88$& 0.65 & $-0.31$ & $<-0.19$& 0.18    & 1.21 & 0.75 \\              
%HE~0317$-$4640 & 5872 &  4.10 & $-2.33$ & 1.49 & $<-0.32$ & $<1.91$ & -    & -       & -       & 0.22    & 1.00 \\              
HE~0323$-$4529 & 5127 &  2.51 & $-3.15$ & 1.62 & $<-1.36$ & $<1.70$ & -    & -       & -       & 0.38    & 0.90 & 0.66 \\               
HE~0328$-$1047 & 5301 &  3.03 & $-2.25$ & 1.21 & $<-1.54$  & $<0.62$& 0.42 & -0.49   & $<-0.23$& 0.15    & 0.88 & 0.25 \\               
%HE~0330$-$4004 & 5946 &  3.78 & $-2.20$ & 1.53 & $<-0.75$ & $<1.36$ & -    & -       & -       & 0.08    & 0.71 \\               
HE~0330$-$4144 & 5961 &  4.20 & $-1.90$ & 1.52 & $<-0.23$ & $<1.58$ & -    & -       & -       & 0.19    & 0.52 & 0.16 \\               
%HE~0331$-$4939 & 5206 &  2.60 & $-2.90$ & 1.55 & $-1.16 $ & 1.65    & -    & -       & -       & 0.34    & 1.26 \\              
%HE~0333$-$4001 & 5892 &  4.31 & $-2.64$ & 1.30 & $-0.05$  & 2.50    & -    & -       & -       & 0.32    & 1.27 \\               
%HE~0337$-$5127 & 5247 &  2.86 & $-2.62$ & 1.55 & $<-1.36$ & $<0.3$  & 0.60 & $-0.62$ & $<0.14$ & 0.16    & 0.53 \\              
%HE~0338$-$3945 & 6162 &  4.09 & $-2.41$ & 1.33 & $<0.07$  & $<2.39$ & 1.89 &  0.52   & $<0.08$ & 2.07    & 1.37 \\      
%HE~0339$-$4027 & 5896 &  4.11 & $-1.81$ & 1.48 & $<-0.60$ & $<1.12$ & -    & -       & -       & 0.18    & 1.08  \\            
%HE~0340$-$3430 & 5914 &  3.81 & $-1.95$ & 1.59 & $<-0.40$ & $<1.02$ & -    & -       & -       & 0.06    & 0.54 \\            
HE~0340$-$5355 & 4862 &  1.81 & $-2.89$ & 1.87 & $<-2.01$  & $<0.79$ & -    & -       & -       & $-0.11$ & 0.90 & 0.52 \\              
HE~0341$-$4024 & 6108 &  4.20 & $-1.82$ & 1.41 & $<-0.48$ & $<1.25$ & 0.69 & $-0.77$ & $<0.14$ & 0.27    & 1.01 & 0.43 \\             
HE~0347$-$1819 & 5198 &  4.23 & $-2.78$ & 1.61 & $<-0.84$ & $<1.85$ & -    & -       & -       & 0.03    & 0.34 & 0.18 \\           
HE~0353$-$6024 & 5320 &  3.15 & $-3.17$ & 1.49 & $-1.06$  & 2.02    & -    & -       & -       & 0.29    & 1.46 & 1.00 \\              
%HE~0417$-$0821 & 5811 &  4.83 & $-2.33$ & 1.40 & $<-0.28$ & $<1.73$ & -    & -       & -       & 0.25    & 0.70 \\             
%HE~0430$-$4404 & 6214 &  4.27 & $-2.07$ & 1.41 & -        & -       & -    & -       & -       & 1.44    &\\             
%HE~0430$-$4901 & 5296 &  3.12 & $-2.72$ & 1.32 & $<-1.11$ & $<1.52$ & 1.16 & $-0.66$ & $-0.06$ & 0.09    & 1.05 \\             
%HE~0432$-$0923 & 5131 &  2.64 & $-3.19$ & 1.54 & -        & -       & 1.25 & $-0.53$ & -       & 0.24    &\\             
%HE~0436$-$4008 & 5431 &  3.34 & $-2.35$ & 1.30 & -        & -       & -    & -       & -       & 0.49    &\\             
%HE~0441$-$4343 & 5629 &  3.59 & $-2.52$ & 1.29 & $-0.53 $ & 1.90    & -    & -       & -       & 0.33    & 1.33  \\  
HE~0442$-$1234 & 4604 & 1.34  & $-2.41$ & 2.21  & $-1.91$ & $0.41$ & 0.52 & $-0.65$ & $-0.53$ & $-0.61$ & 5.18   & 2.16 \\        
%HE~0447$-$4858 & 5995 &  3.86 & $-1.69$ & 1.35 & -        &  -      & -    & -       & -       & 0.04    &\\             
HE~0450$-$4705 & 5429 &  3.34 & $-3.10$ & 1.49 & $<-1.00$ & $<2.01$ & -    & -       & -       & 0.84    & 0.53 & 0.17 \\             
%HE~0454$-$4758 & 5388 &  3.33 & $-3.10$ & 1.49 & -        & -       & -    & -       & -       & 0.44    &\\              
HE~0501$-$5139 & 5861 &  3.54 & $-2.38$ & 1.49 & $< 0.11$ & $<2.4$ & -    & -       & -       & 0.40    & 0.20 & 0.17 \\              
%HE~0501$-$5644 & 5033 &  2.30 & $-2.41$ & 1.76 & $-1.42$  & 0.90    & 0.30 & $-0.75$ &  0.17   & 0.27    & 1.48\\              
%HE~0513$-$4557 & 5629 &  3.62 & $-2.79$ & 1.44 & -        & -       & -    &  -      & -       & 0.39    &   \\             
%HE~0516$-$3820 & 5342 &  3.05 & $-2.33$ & 1.48 & -        & -       & 0.67 & $-0.78$ & -       & 0.39    &\\               
HE~0520$-$1748 & 5272 &  3.06 & $-2.52$ & 1.46 & $<-1.82$  & $<0.61$ & -    & -       & -       & 0.45    & 1.59 & 0.92 \\              
HE~0524$-$2055 & 4739 &  1.57 & $-2.58$ & 1.95 & $-1.91$  & $0.58$ & 0.49 & $-0.42$ & $-0.32$ & $-0.25$ & 2.48 & 1.34 \\              
HE~0534$-$4615 & 5506 &  3.40 & $-2.01$ & 1.43 & $<-1.15$ & $<0.77$ & 0.49 & $-0.50$ & $<-0.15$& 0.13    & 1.05 & 0.28 \\              
%HE~0547$-$4539 & 5152 &  2.59 & $-3.01$ & 1.57 & $<-1.78$ & $<0.94$ & -    & -       & -       & 0.50    & 0.83 \\               
HE~0926$-$0508 & 6249 &  4.24 & $-2.78$ & 1.60 & $<-0.40$ & $<2.29$ & -    & -       &-        & 0.62    & 0.21 & 0.12 \\              
%HE~0938$+$0114 & 6777 &  4.89 & $-2.51$ & 1.76 & $<-0.13$ & $<2.07$ & -    & -       & -       & 0.65    & 0.68 \\               
%HE~0951$-$1152 & 5349 &  4.75 & $-2.62$ & 1.42 & $<-0.41$ & $<1.82$ & -    & -       & -       & 0.10    & 0.67 \\               
%HE~1015$-$0027 & 6315 &  4.40 & $-2.66$ & 1.43 & $<0.23$  & $<1.55$ & -    & -       & -       & -       & 0.09 \\              
HE~1052$-$2548 & 6534 &  4.52 & $-2.29$ & 1.57 & $< 0.06$ & $<2.26$ & -    & -       & -       & 0.51    & 0.73 & 0.63 \\               
%HE~1059$-$0118 & 5560 &  4.30 & $-2.81$ & 1.28 & $<-0.52$ & $<1.92$ & -    & -       & -       & 0.37    & 0.67 \\              
HE~1100$-$0137 & 6101 &  4.25 & $-2.92$ & 1.29 & $<-0.02$ & $<2.81$ & -    & -       & -       & 0.47    & 0.17 & 0.17 \\              
%HE~1105$+$0027 & 6132 &  3.45 & $-2.42$ & 1.52 & $<-0.07$ & $<2.26$ & 1.81 &  0.64   &  0.03   & 2.00    & 1.20 \\              
%HE~1120$-$0153 & 6191 &  4.09 & $-2.77$ & 0.68 & $<-0.57$ & -       & -    & -       & -       & 0.63    & \\              
%HE~1122$-$1429 & 5787 &  3.29 & $-2.65$ & 1.54 & $<-0.58$ & $<1.34$ & -    & -       & -       & 0.44    & 0.45 \\              
HE~1124$-$2335 & 5226 &  2.68 & $-2.95$ & 1.65 & $<-1.41$ & $<1.45$ & -    & -       & -       & 0.86    & 0.53 & 0.11 \\              
HE~1126$-$1735 & 5689 &  3.31 & $-2.69$ & 1.55 & $<-0.84$ & $<1.76$ & -    & -       & -       & 0.23    & 0.68 & 0.15 \\               
HE~1127$-$1143 & 5224 &  2.64 & $-2.73$ & 1.59 & $<-1.31$ & $<1.33$ & 1.08 & $-0.45$ & $<-0.17$ & 0.54    & 1.10 & 0.76 \\              
%HE~1131$+$0141 & 5236 &  2.98 & $-2.48$ & 1.63 & $<-1.10$ & $<1.03$ & 0.87 & $-0.35$ & $<0.00$ & 0.12    & 0.68 \\              
%HE~1132$+$0125 & 5732 &  3.54 & $-2.42$ & 1.41 & $<-0.34$ & $<1.99$ & -    & -       & -       & 0.24    & 1.83 \\              
HE~1132$+$0204 & 5046 &  2.25 & $-2.55$ & 1.68 & $<-1.65$  & $<0.81$& 0.25 & $-0.94$ &  $<0.12$& 0.13    & 1.41 & 0.95 \\              
%HE~1135$-$0344 & 6154 &  4.03 & $-2.63$ & 1.17 & $<0.24$  & $<2.78$ & -    &  -      & -       & 1.03    & 1.51 \\              
%HE~1148$-$0037 & 5964 &  4.16 & $-3.47$ & 0.81 & $<-0.28$ & $<3.10$ & -    &  -      & -       & 0.84    & 1.05 \\              
HE~1207$-$2031 & 6281 &  4.40 & $-2.82$ & 1.42 & $<0.00$ & $<2.73$ & -    & -       & -       & 0.64    & 0.94 & 0.69 \\              
%HE~1210$+$0048 & 6028 &  3.73 & $-2.28$ & 1.56 & -        & -       & -    & -       & -       & 0.57    &\\              
%HE~1210$-$1956 & 5790 &  3.30 & $-2.57$ & 1.52 & -        & -       & -    & -       & -       & 0.22    &\\              
%HE~1212$-$0127 & 4915 &  1.85 & $-2.15$ & 1.87 & -        & -       & 0.35 & $-0.67$ & -       & $-0.39$ & \\              
%HE~1214$-$1819 & 4916 &  1.88 & $-3.01$ & 1.74 & $<-2.02$ & $<0.90$ & 0.49 & $-0.53$ & $<-0.02$& 0.35    & 1.13\\              
%HE~1215$+$0149 & 5098 &  2.37 & $-2.90$ & 1.48 & -        & -       & -    & -       & -       & 0.15    &\\              
%HE~1217$-$0540 & 5700 &  4.20 & $-2.95$ & 1.41 & $<-0.02$ & $<2.84$ & -    & -       & -       & 0.81    & 1.79\\              
%HE~1219$-$0312 & 5140 &  2.40 & $-2.81$ & 1.54 & -        & -       & 1.41 & $-0.90$ & -       & $-0.08$ &  \\      
%HE~1221$-$0522 & 5789 &  4.04 & $-2.84$ & 1.29 & $<-0.28$ & $<1.76$ &  -   & -       & -       & 0.53    & 0.24 \\             
HE~1221$-$1948 & 6083 &  3.81 & $-2.60$ & 1.65 & $<-0.26$  & $<3.01$  & -   & -       & -        & 1.42    & 1.46 & 1.36 \\              
%HE~1222$-$0200 & 5288 &  2.72 & $-2.45$ & 1.50 & -        & -       & 0.60 & $-0.60$ &  0.53   & 0.23    & \\               
%HE~1222$-$0336 & 6052 &  3.96 & $-2.04$ & 1.30 & $-0.11$  &  1.84   & -    & -       & -       & 0.22    & 1.05 \\              
HE~1225$+$0155 & 4842 &  1.80 & $-2.75$ & 1.85 & $<-2.50$  & $<0.16$& 0.25 & $-0.70$ & $<-0.51$ & 0.26    & 0.90 & 0.23 \\               
%HE~1230$-$1724 & 5952 &  3.98 & $-2.30$ & 1.39 & $<-0.27$ & $<1.09$ & -    & -       & -       & 0.13    & 0.24 \\              
%HE~1243$-$1425 & 5507 &  3.19 & $-2.67$ & 1.11 & $ -0.65$ & 1.93    & -    & -       & -       & 0.51    & 1.63\\               
HE~1245$-$1616 & 6191 &  4.04 & $-2.98$ & 1.53 & $<-0.03$  & $<2.86$ & -    & -       & -       & 0.77    & 0.79 & 0.73 \\               
HE~1246$-$1344 & 4853 &  1.65 & $-3.40$ & 1.84 & $<-2.16$  & $<1.15$& -    & -       & -       & $-0.06$ & 0.89 & 0.30 \\               
HE~1247$-$2114 & 5012 &  2.08 & $-2.61$ & 1.67 & $<-1.87$  & $<0.65$ & 0.22 & $-0.65$ & $<0.01$& 0.32    & 1.26 & 0.66 \\              
%HE~1248$-$1800 & 5288 &  2.92 & $-2.89$ & 1.47 & -        & -       & -    & -       & -       &  0.53   & \\               
%HE~1249$-$2932 & 4718 &  1.52 & $-2.65$ & 2.09 & -        & -       & -    & -       & -       & $-0.41$ & \\              
%HE~1249$-$3121 & 5373 &  3.40 & $-3.23$ & 1.58 & $<-0.78$ & $<1.98$ & -    & -       & -       & 1.86    & 0.49 \\                
HE~1251$-$0104 & 5084 &  2.32 & $-2.73$ & 1.58 & $<-1.64$ & $<1.00$ & -    & -       & -       & 0.25    & 0.83 & 0.46 \\               
%HE~1252$+$0044 & 5296 &  2.98 & $-3.28$ & 1.57 & $<-1.06$ & $<2.09$ & -    & -       & -       & 0.60    & 0.95 \\                
%HE~1252$-$0117 & 4847 &  1.67 & $-2.89$ & 1.88 & -        & -       & -    & -       & -       & $-0.16$ &\\               
%HE~1254$+$0009 & 4865 &  1.86 & $-2.94$ & 1.78 & -        & -       & -    & -       & -       & $-0.11$ &\\               
HE~1256$-$0651 & 6137 &  4.05 & $-2.36$ & 1.50 & $<-0.23$  & $<2.04$ & -    & -       & -       & 0.62    & 0.93 & 0.45 \\               
%HE~1259$-$0621 & 5787 &  3.68 & $-2.64$ & 1.60 & $<-0.43$ & $<1.81$ & -    & -       & -       & 0.41    & 0.61 \\               
%HE~1300$+$0157 & 5411 &  3.38 & $-3.76$ & 1.43 & $<-0.87$ & $<2.44$ & -    & -       & -       & 1.17    & 0.51 \\               
%HE~1300$-$0641 & 5308 &  2.96 & $-3.14$ & 1.59 & $<-0.98$ & $<1.26$ & -    & -       & -       & 1.29    & 0.22 \\                
HE~1300$-$0642 & 5173 &  2.68 & $-3.03$ & 1.57 & $<-1.25$ & $<1.69$ & -    & -       & -       & 0.34    & 0.49 & 0.24 \\               
%HE~1300$-$2201 & 6332 &  4.64 & $-2.61$ & 1.34 & -        & -       & -    & -       & -       & 1.01    &\\               
HE~1300$-$2431 & 5029 &  1.96 & $-3.25$ & 1.92 & $<-1.82$ & $<1.34$ & -    & -       & -       & $-0.16$ & 0.95 & 0.47 \\               
HE~1305$-$0331 & 6081 &  4.22 & $-3.26$ & 1.58 & $<-0.28$ & $<2.89$ & -    & -       & -       & 1.13    & 0.78 & 0.29 \\  
%HE~1311$-$1412 & 4796 &  1.50 & $-2.91$ & 1.92 & $-1.80$  & 1.02    & 0.53 & $-0.59$ & 0.04    & $-0.15$ & 1.86 \\            
%HE~1314$-$3036 & 4757 &  1.54 & $-2.99$ & 1.88 & -        &-        & 0.17 & $-0.22$ & -       & $-0.13$ &\\               
%HE~1320$-$1339 & 4935 &  1.69 & $-2.78$ & 1.97 & $<-2.26$ & $<0.43$ & 0.16 & $-0.58$ & $<-0.15$& $-0.51$ & 3.52 \\               
HE~1330$-$0354 & 6257 &  4.13 & $-2.29$ & 1.49 & $<0.04$ & $<2.24$ &  -   & -       & -       & 1.05    & 0.54 & 0.31 \\               
%HE~1330$-$0607 & 5094 &  2.30 & $-2.33$ & 1.90 & $<-1.49$ & $<0.73$ & 0.25 & $-1.01$ & $<0.08$ & 0.21    & 0.99 \\               
HE~1332$-$0309 & 5125 &  2.40 & $-2.46$ & 1.64 & $-0.97 $ & $1.40$ & 0.48 & $-0.48$ & $0.50$ & 0.21    & 1.96 & 1.43 \\               
%HE~1333$-$0340 & 6053 &  4.18 & $-2.64$ & 1.37 & $<0.07$  & $<2.28$ & -    & -       & -       & -       & 0.52 \\              
%HE~1337$+$0012 & 6141 &  4.25 & $-3.44$ & 1.49 & $<-0.67$ & $<0.66$ & -    & -       & -       & 0.71    & 0.11 \\               
HE~1337$-$0453 & 5938 &  3.56 & $-2.34$ & 1.62 & $<-0.41$ & $<1.84$ & -    & -       & -       & 0.12    & 0.46 & 0.23 \\               
%HE~1343$-$0640 & 5942 &  3.97 & $-1.90$ & 1.48 & $<-0.16$ & $<0.45$ & -    & -       & -       & 0.77    & 0.26 \\               
%HE~1345$-$0206 & 5006 &  2.22 & $-2.82$ & 1.65 & $<-1.44$ & $<1.18$ & -    & -       & -       & 0.34    & 0.85 \\                
%HE~1413$-$1954 & 6533 &  4.59 & $-3.22$ & 1.40 & $< 0.41$ & $<3.06$ & -    & -       & -       & 1.45    & 0.37 \\               
%HE~1419$-$1759 & 4809 &  1.63 & $-3.18$ & 1.83 & -        & -       & -    & -       & -       & $-0.20$ &\\               
%HE~1421$-$2006 & 5687 &  3.73 & $-2.65$ & 1.33 & $<-0.52$ & $<1.29$ & -    & -       & -       & 0.30    & 0.26 \\               
%HE~1430$+$0053 & 5201 &  2.74 & $-3.03$ & 1.55 & $<-1.49$ & $<0.38$ & 0.72 & $-0.82$ & $<0.31$ & 0.29    & 0.37 \\               
%HE~1430$-$0026 & 5855 &  4.12 & $-2.79$ & 1.38 & $<-0.31$ & $<2.37$ & -    & -       & -       & 0.52    & 1.00 \\               
%HE~1430$-$1123 & 5915 &  3.75 & $-2.71$ & 1.40 & $<-0.11$ & $<2.51$ & -    & -       &-        & 1.84    & 1.23\\               
HE~1431$-$2142 & 6137 &  4.10 & $-2.60$ & 1.55 & $<-0.20$ & $<2.31$ & -    & -       & -       & 0.48    & 0.60 & 0.23 \\              
%HE~1500$-$1628 & 4994 &  2.10 & $-2.31$ & 1.85 & -        & -       & -    & -       & -       & 0.13    &\\               
HE~2133$-$1432 & 5716 &  3.46 & $-2.02$ & 1.46 & $<-0.82$  & $<1.11$ & -    & -       & -       & 0.12    & 1.10 & 0.60 \\               
HE~2134$+$0001 & 5257 &  3.00 & $-2.22$ & 1.55 & $<-1.36$  & $<0.77$ & 0.47 & $-0.89$ & $<-0.12$ & 0.20    & 1.17 & 0.52 \\               
%HE~2139$-$1851 & 4925 &  1.86 & $-3.25$ & 1.78 & $<-2.02$ & $<0.86$ & -    & -       & -       & 0.49    & 0.61 \\               
%HE~2150$-$0825 & 5960 &  3.67 & $-1.98$ & 1.57 & $<-0.18$ & $<1.71$ & -    & -       & -       & 1.35    & 1.14\\               
HE~2151$-$2858 & 5598 &  4.14 & $-2.38$ & 1.39 & $<-0.43$ & $<1.86$ & -    & -       & -       & 0.10    & 0.68 & 0.51 \\               
HE~2153$-$2719 & 4898 &  2.01 & $-2.49$ & 1.80 & $<-2.81$ & $<-0.41$ & 0.21 & $-0.90$ & $<-1.04$& 0.12    & 0.95 & 0.24\\               
%HE~2154$-$2838 & 5303 &  3.18 & $-1.85$ & 1.46 & -        & -       & 0.45 & $-0.48$ & -       & 0.05    &\\                
%HE~2155$+$0136 & 5331 &  3.22 & $-2.07$ & 1.23 & $<-0.78$ & $<1.20$ & 0.68 & $-0.55$ & $<0.09$ & 0.00    & 1.32\\                
%HE~2156$-$3130 & 4692 &  1.28 & $-3.13$ & 1.93 & $-1.86$  &  1.18   & -    & -       & -       & 0.74    & 2.49\\                
HE~2158$-$3112 & 4843 &  1.85 & $-2.75$ & 1.85 & $<-2.81$ & $<-0.15$ & 0.02 & $-0.86$ & $<0.60$ & $-0.04$ & 1.31 & 0.42 \\               
%HE~2201$-$0637 & 4976 &  2.21 & $-2.61$ & 1.69 & $<-1.91$ & $<0.17$ & -    & -       & -       & 0.14    & 0.76 \\               
%HE~2216$-$0621 & 4671 &  1.27 & $-3.23$ & 2.02 & -        & -       & -    & -       & -       & $-0.66$ &\\               
%HE~2217$-$1523 & 4847 &  1.82 & $-2.62$ & 1.81 & $<-2.08$ & $<0.37$ & -    & -       & -       & 0.04    & 0.93 \\                
HE~2219$-$0713 & 4789 &  1.68 & $-2.91$ & 1.64 & $-1.97$  & 0.85    & -    & -       & -       & $-0.17$ & 1.35 & 0.97 \\                
%HE~2221$-$4150 & 5887 &  4.06 & $-2.03$ & 1.41 & $<-0.38$ & $<1.21$ & -    & -       & -       &  0.23   & 0.59 \\               
%HE~2222$-$4156 & 5537 &  3.27 & $-2.73$ & 1.44 & -        & -       & -    & -       & -       &  0.42   &\\               
HE~2224$+$0143 & 5198 &  2.66 & $-2.58$ & 1.67 & $-1.47$  & 1.02    & 1.05 & $-0.46$ & $-0.45$ &  0.35   & 1.72 & 0.99 \\               
HE~2224$-$4103 & 5074 &  2.32 & $-2.64$ & 1.75 & $<-1.71$ & $<0.84$ & -    & -       & -       &  0.23   & 0.90 & 0.45 \\               
%HE~2226$-$4102 & 5140 &  2.43 & $-2.87$ & 1.73 & $<-1.40$ & $<1.01$ & -    & -       & -       &  0.46   & 0.59 \\                
%HE~2227$-$4044 & 5811 &  3.85 & $-2.32$ & 1.40 & $<-0.34$ & $<1.87$ & -    & -       & -       &  1.67   & 0.97 \\                
%HE~2228$-$3806 & 5175 &  2.62 & $-3.07$ & 1.65 & $<-1.40$ & $<0.72$ & -    & -       & -       &  0.42   & 0.39 \\                
HE~2229$-$4153 & 5138 &  2.47 & $-2.62$ & 1.79 & $<-1.91$ & $<0.62$ & 0.45 & $-0.73$ & $<-0.28$&  0.37   & 0.79 & 0.20 \\               
%HE~2231$-$0622 & 5211 &  2.90 & $-2.12$ & 1.50 & $<-0.85$ & $<1.03$ & -    & -       & -       & $-0.08$ & 0.82 \\               
HE~2234$-$0521 & 5332 &  3.15 & $-2.78$ & 1.39 & $<-0.93$ & $<1.76$ & -    & -       & -       &  0.36   & 0.27 & 0.07 \\               
%HE~2238$-$2152 & 5427 &  3.28 & $-2.40$ & 1.46 & $<-0.99$ & $<1.17$ & -    & -       & -       &  0.13   & 0.78 \\                
%HE~2240$-$0412 & 5852 &  4.33 & $-2.20$ & 1.30 & $<0.02$  & $<1.71$ & -    & -       & -       &  1.35   & 0.46 \\                
HE~2247$-$3705 & 5366 &  3.04 & $-2.27$ & 1.48 & $<-1.39$ & $<0.79$ & -    & -       & -       &  0.36   & 0.76 & 0.27 \\               
%HE~2248$-$3345 & 5011 &  2.11 & $-2.74$ & 1.71 & -        & -       & -    & -       & -       &  0.21   &\\               
HE~2250$-$2132 & 5705 &  3.69 & $-2.22$ & 1.43 & $<-0.74$ & $<1.39$ & -    & -       & -       & 0.41    & 0.89 & 0.43 \\               
HE~2252$-$4157 & 5090 &  2.87 & $-1.93$ & 1.55 & $-1.32$  & 0.52    & 0.53 & $-0.24$ & $-0.43$ & $-0.15$ & 2.98 & 1.34 \\              
HE~2259$-$3407 & 6266 &  4.32 & $-2.29$ & 1.37 & $<0.00$  & $<2.2$  & -    & -       & -       &  0.41   & 0.39 & 0.16 \\               
HE~2311$+$0129 & 5188 &  2.65 & $-2.78$ & 1.54 & $-1.32$  & 1.37    & -    & -       & -       &  0.33   & 1.24 & 1.05 \\               
%HE~2325$-$0755 & 5665 &  3.17 & $-2.85$ & 1.48 & $<-0.76$ & $<1.84$ & -    & -       & -       &  0.21   & 0.74 \\               
HE~2326$+$0038 & 5145 &  2.51 & $-2.77$ & 1.62 & $<-1.59$ & $<1.09$ & -    & -       & -       &  0.23   & 0.51 & 0.11 \\               
HE~2327$-$5642 & 5048 &  2.22 & $-2.95$ & 1.69 & $-1.45$  & 1.41    & 1.22 & $-0.56$ & $-0.25$ &  0.43   & 1.79 & 1.35 \\       
%HE~2329$-$3702 & 6060 &  3.72 & $-2.16$ & 1.65 & $<-0.13$ & $<0.01$ &  -   & -       & -       &  0.15   & 0.12 \\               
%HE~2333$-$1358 & 5020 &  2.15 & $-3.34$ & 1.57 & -        & -       &  -   & -       & -       &  0.33   &\\ 
%HE~2338$-$1311 & 5582 &  3.50 & $-2.86$ & 1.27 & -        & -       &  -   & -       & -       & 0.34    & \\               
HE~2338$-$1618 & 5515 &  3.38 & $-2.65$ & 1.43 & $<-0.76$ & $<1.80$ &  -   & -       & -       &  0.47   & 0.88 & 0.56 \\               
HE~2345$-$1919 & 5617 &  4.46 & $-2.46$ & 1.47 & $<-0.47$  & 1.90   &  -   & -       & -       & 0.24    & 0.97 & 0.60 \\               
%HE~2347$-$1254 & 6132 &  3.95 & $-1.83$ & 1.67 & $<-0.45$ & $<0.92$ &  -   & -       & -       & 0.27    & 0.66 \\               
%\hline
\end{longtable}
%\end{landscape}
%}% End onllongtab

\begin{table*}
\centering
\caption{Thorium and europium abundances from previous works \citep[From the summary of SAGA database and][]{roederer:09}. Both detected values and upper limits are included.}\label{tab:otherworks}

\begin{tabular}{l r r r r r c}
\hline\hline
        name &[Fe/H] & log$\epsilon$(Th)&log$\epsilon$(Eu)& log(Th/Eu)& "Age"& Ref. \\ 
             &       &                  &                 &           & [Gyr]&       \\       
\hline
        BD+08\_2856      & $-2.10$ & $-1.78\pm0.10$ & $-1.16\pm0.06$ & $-0.62$ & 13.5    & 1,2 \\
        BD+17\_3248      & $-2.23$ & $-1.18\pm0.10$ & $-0.67\pm0.05$ & $-0.51$ &  8.4    & 3 \\
        BD+04\_2621      & $-2.50$ & $<-3.12$       & $-2.61\pm0.23$ & $<-0.51$& $>8.4$  & 1,2  \\
        BD$-$18\_5550    & $-3.03$ & $<-3.10$       & $-2.79\pm0.22$ & $<-0.31$& $>-0.9$ & 1,2 \\
        CS22892$-$052    & $-2.92$ & $-1.42\pm0.15$ & $-0.86\pm0.02$ & $-0.56$ & 10.7    & 7 \\
        CS29491$-$069    & $-2.60$ & $-1.43\pm0.22$ & $-1.03\pm0.10$ & $-0.40$ &  3.3    & 3 \\
        CS29497$-$004    & $-2.81$ & $-0.96\pm0.15$ & $-0.45\pm0.20$ & $-0.28$ & $-2.3$  & 4,5\\
        CS30306$-$132    & $-2.40$ & $-1.12\pm0.15$ & $-1.14\pm0.25$ & $ 0.02$ & $-16.3$ & 7 \\
        CS31078$-$018    & $-2.84$ & $-1.35\pm0.25$ & $-1.17\pm0.17$ & $-0.18$ & $-7.0$  & 8 \\
        CS31082$-$001    & $-2.90$ & $-0.98\pm0.05$ & $-0.76\pm0.11$ & $-0.22$ & $-5.1$  & 8,9,10 \\
        HD108317         & $-2.18$ & $-1.84\pm0.20$ & $-1.32\pm0.05$ & $-0.52$ &  8.9    & 3 \\
        HD108577         & $-2.36$ & $-2.11\pm0.14$ & $-1.48\pm0.02$ & $-0.63$ & 14.0    & 1,2 \\
        HD110184         & $-2.52$ & $-2.50\pm0.15$ & $-1.91\pm0.05$ & $-0.59$ & 12.1    & 7 \\
        HD115444         & $-2.85$ & $-1.97\pm0.15$ & $-1.64\pm0.03$ & $-0.33$ &  0.0    & 3 \\
        HD122563         & $-2.72$ & $<-2.43$       & $-2.75\pm0.11$ & $<0.32$& $>-30.3$ & 3  \\
        HD122956         & $-1.95$ & $-1.50\pm0.17$ & $-0.94\pm0.07$ & $-0.56$ & 10.7    & 3 \\
        HD126587         & $-2.93$ & $<-2.39$       & $-1.97\pm0.06$ & $<-0.42$& $>4.2$  & 3 \\
        HD128279         & $-2.00$ & $<-2.00$       & $-1.57\pm0.06$ & $<-0.43$& $>4.7$  & 2,18\\
        HD175305         & $-1.48$ & $-0.76\pm0.15$ & $-0.36\pm0.07$ & $-0.40$ &  3.3    & 3 \\
        HD186478         & $-2.50$ & $-1.85\pm0.15$ & $-1.34\pm0.06$ & $-0.51$ &  8.4    & 7 \\
        HD204543         & $-1.87$ & $-1.68\pm0.14$ & $-1.05\pm0.07$ & $-0.63$ & 14.0    & 6 \\
        HD221170         & $-2.20$ & $-1.46\pm0.05$ & $-0.86\pm0.07$ & $-0.60$ & 12.6    & 11 \\
        HD6268           & $-2.40$ & $-1.93\pm0.10$ & $-1.56\pm0.03$ & $-0.37$ &  1.9    & 3 \\
        HD74462          & $-1.52$ & $-0.94\pm0.13$ & $-0.50\pm0.09$ & $-0.44$ &  5.1    & 3 \\
        HD88609          & $-3.07$ & $<-2.65$       & $-2.89\pm0.12$ & $<0.24 $& $>-27$  & 19 \\
        HE0338$-$3945    & $-2.42$ & $<0.23$        & $ 0.02\pm0.17$ & $<0.21 $& $>-25$  & 6 \\
        HE1219$-$0312    & $-2.97$ & $-1.29\pm0.14$ & $-1.06\pm0.10$ & $-0.23$ & $-4.7$  & 3 \\
        HE1523$-$0901    & $-2.95$ & $-1.20\pm0.05$ & $-0.62\pm0.05$ & $-0.58$ & 11.7    & 12 \\
        HE2148$-$1247    & $-2.50$ & $<-0.50$       & $ 0.17\pm0.10$ & $<-0.67$& $>15.9$ & 20 \\
        HE2327$-$5642    & $-2.78$ & $-1.67\pm0.21$ & $-1.29\pm0.07$ & $-0.38$ &  2.3    & 13\\ 
        M15 K341         & $-2.32$ & $-1.51\pm0.10$ & $-0.88\pm0.09$ & $-0.63$ & 14.0    & 14 \\
        M15 K462         & $-2.25$ & $-1.30\pm0.10$ & $-0.61\pm0.09$ & $-0.69$ & 16.8    & 14 \\
        M15 K583         & $-2.34$ & $-1.70\pm0.10$ & $-1.24\pm0.09$ & $-0.46$ &  6.1    & 14 \\
        M5 IV$-$81       & $-1.28$ & $-0.58\pm0.15$ & $-0.31\pm0.05$ & $-0.27$ & $-2.8$  & 15,16 \\
        M5 IV$-$82       & $-1.33$ & $-0.68\pm0.15$ & $-0.23\pm0.05$ & $-0.45$ &  5.6    & 15,16 \\
        M92 VII$-$18     & $-2.29$ & $-2.01\pm0.07$ & $-1.45\pm0.07$ & $-0.56$ & 10.7    & 1,2 \\
        UMi COS82        & $-1.42$ & $-0.25\pm0.15$ & $ 0.34\pm0.11$ & $-0.59$ & 12.1    & 17 \\
        
\hline
\end{tabular}
\tablebib{(1)~\citet{johnson:02}; (2) \citet{johnson:01}; (3) \citet{roederer:09}; (4) \citet{christlieb:04}; (5) \citet{barklem:05}; (6) \citet{jonsell:06}; (7) \citet{honda:04}; (8) \citet{hill:02}; (9) \citet{plez:04}; (10) \citet{sneden:09}; (11) \citet{ivans:06}; (12) \citet{frebel:07}; (13) \citet{mashonkina:10}; (14) \citet{sneden:00}; (15) \citet{yong:08b}; (16) \citet{yong:08a}; (17) \citet{aoki:07}; (18) \citet{simmerer:04}; (19) \citet{honda:07}; (20) \citet{cohen:03}.
}
\end{table*}

\end{document}